\documentclass{jfm}
\usepackage{graphicx}
\usepackage{epstopdf, epsfig}
\usepackage{xcolor}
\usepackage{amsmath}
\usepackage[permil]{overpic}
\usepackage{hyperref}

\usepackage{gensymb}
\usepackage{algorithm}
\usepackage{algpseudocode}
\usepackage{soul}
\usepackage{adjustbox}
\usepackage{tabularx}
\usepackage[normalem]{ulem}

\usepackage{newtxtext}
\usepackage{newtxmath}
\usepackage{xfrac}					
\usepackage{tikz}
\usepackage{bm}

\shorttitle{Neural networks in feedback for flow analysis, sensor placement and control}
\shortauthor{T. C. Déda, W. R. Wolf and S. T. M. Dawson}
\title{Neural networks in feedback for flow analysis, sensor placement and control}

\author{Tarcísio C. Déda\aff{1}
  \corresp{\email{tarcisio.deda@gmail.com}},
  William R. Wolf\aff{1}
 \and Scott T. M. Dawson\aff{2}}

\affiliation{\aff{1}School of Mechanical Engineering, University of Campinas, R. Mendeleyev 200, Campinas, São Paulo, Brazil, 13083-860
\aff{2}Mechanical, Materials, and Aerospace Engineering Department, Illinois Institute of Technology, 10 W 35th St, Chicago, IL, United States, 60616}

\begin{document}

\maketitle

\begin{abstract}

This work presents a novel methodology for analysis and control of nonlinear fluid systems using neural networks. The approach is demonstrated on four different study cases being the Lorenz system, a modified version of the Kuramoto-Sivashinsky equation, a streamwise-periodic 2D channel flow, and a confined cylinder flow. Neural networks are trained as models to capture the complex system dynamics and estimate equilibrium points through a Newton method, enabled by backpropagation. These neural network surrogate models (NNSMs) are leveraged to train a second neural network, which is designed to act as a stabilizing closed-loop controller. The training process employs a recurrent approach, whereby the NNSM and the neural network controller (NNC) are chained in closed loop along a finite time horizon. By cycling through phases of combined random open-loop actuation and closed-loop control, an iterative training process is introduced to overcome the lack of data near equilibrium points. This approach improves the accuracy of the models in the most critical region for achieving stabilization. Through the use of L1 regularization within loss functions, the NNSMs can also guide optimal sensor placement, reducing the number of sensors from an initial candidate set. The datasets produced during the iterative training process are also leveraged for conducting a linear stability analysis through a modified dynamic mode decomposition approach. The results demonstrate the effectiveness of computationally inexpensive neural networks in modeling, controlling, and enabling stability analysis of nonlinear systems, providing insights into the system behaviour and offering potential for stabilization of complex fluid systems.
\end{abstract}


\section{Introduction}

Neural networks (NNs) are powerful tools for data-driven mathematical modelling, finding a range of applications in the field of fluid mechanics such as flow analysis, optimization and control \citep{gadelhak1996modern, lee1997application, Kutz2017, brunton2020machine}. Recent advances in experimental data sampling, the availability of data from increasingly complex simulations, and the widespread availability of libraries that facilitate the design and training of NNs has encouraged their rapid development and application for the purposes of data compression, model order reduction, flow analysis, turbulence modelling, and feature extraction, all leveraging the power of NNs as general nonlinear function approximators. 
In the field of turbulence modeling, deep neural networks have been used to improve the prediction capabilities of RANS closure models \citep{ling2016reynolds,maulik2019subgrid}. 
In the area of flow estimation and reconstruction, \citet{milano2002neural} presented one of the first uses of neural networks in fluid mechanics for reconstructing near-wall turbulent flow fields. More recently, for example, \citet{fukami2019} employed convolutional neural networks (CNNs) to accurately reconstruct turbulent flows from coarse flowfield images. In the context of reduced order modeling, \citet{lui_wolf_2019} combined deep neural networks and proper orthogonal decomposition (POD) to build accurate models of unsteady flows involving transient features, rewriting the Navier-Stokes equations as a set of nonlinear ordinary differential equations. More recently, \cite{miotto_pof_2023} employed vision transformers and CNNs to extract quantities of interest such as aerodynamic coefficients and skin friction from flowfield visualizations. These authors demonstrated that the neural network models based on input images were effective in interpolating and extrapolating between flow regimes, offering a promising alternative for sensors in experimental campaigns and for building surrogate models of complex unsteady flows.

In the field of flow control, different techniques based on training NNs are found in the literature to either build a model for enabling control design \citep{morton2018deep,bieker2020deep} or to directly infer a suitable control law for a given task. Training of NN-based models for control design was conducted by \citet{bieker2020deep}, who applied limited sensor data supported by delay coordinates, and \citet{morton2018deep}, in an approach that modelled the dynamics of a cylinder flow with models trained to learn state mappings to span a Koopman invariant subspace. 
Regarding model-free approaches, reinforcement learning (RL) can be highlighted as a promising tool for automatic training of control strategies with applications to drag reduction, for example. In their work, \citet{rabault2019artificial}, \citet{ren2021applying} and \citet{castellanos2022machine} developed RL strategies to train controllers able to reduce drag in a confined cylinder flow through a set of measurements from velocity probes.  Their sensor/actuation setups are employed to test the techniques proposed in the current work. Similarly, \citet{li2022reinforcement} considered RL strategies to suppress vortex shedding utilising stability analyses and unstable equilibrium computation of the underlying system to obtain improved results. 
Approaches based on RL have also been successfully applied to complex turbulent flows in both experimental \citep{fan2020reinforcement} and computational \citep{guastoni2023deep} setups, being able, in some cases, to stabilise the flows \citep{sonoda2023reinforcement}. Comprehensive reviews of RL for flow control can be found in \citet{vignon2023recent} and \citet{viquerat2022review}.

Models based on sparse identification of nonlinear dynamics (SINDy) \citep{Brunton2016} have been combined with order reduction techniques, such as POD, as an alternative to black box NN models, being able to find simpler and more intuitive models based on a library of interpretable functions. The approach can be applied for building nonlinear models with control inputs, which can be leveraged for closed-loop control applications \citep{kaiser2018sparse}.
On the topic of flow optimization through open-loop control, genetic algorithms can be a candidate tool for parameter search. In such case, studies have been performed to find the best active control setup to improve a fitness function that aims to reduce drag in a bluff body \citep{zigunov2022bluff}, to maximise the thrust vectoring angle in a supersonic jet \citep{zigunov2023multiaxis}, and to find the best open-loop rotation values in order to reduce drag and symmetrise the wake of the flow past a cluster of cylinders \citep{raibaudo2020machine}. 

Traditional feedback control design approaches often rely on well established control theory, whose literature provides techniques that can ensure optimality, robustness, adaptability or the achievement of specific performance characteristics such as settling (convergence) time. 
Such linear control methods have been applied for the control of a variety of flow problems \citep{sipp2016linear}, such as cavities \citep{rowley2006dynamics,barbagallo2009closed}, bluff body wakes \citep{illingworth2016model,flinois2016feedback}, and unsteady aerodynamic systems \citep{brunton2014state,herrmann2022gust,sedky2023experimental}.
However, the broad application of such approaches in fluid mechanics is limited by the fact that they typically rely on linear approximations of the dynamics near desired equilibrium points. Furthermore, building data-driven linear models of flows is also hindered by the fact that signals sampled from fluid flows can be governed by nonlinear phenomena such as chaos, limit cycles, multiple equilibria, switching, and hysteresis. In the case of unstable flows (more specifically globally unstable flows), datasets containing measurements at (approximately) linear regions near equilibrium can be totally absent. Alternatively, nonlinear control techniques, which include sliding-mode control \citep{shi2019global, baek2019adaptive}, feedback linearization \citep{alyoussef2019review}, backstepping control \citep{zulu2016review}, extremum seeking control \citep{deda2021extremum} and switched control \citep{egidio2022trajectory}, can be applied to the task of controlling nonlinear systems, although stabilization relative to a natural equilibrium point may still require a suitable model with some level of accuracy near equilibrium. Differentiable nonlinear models built from data can be leveraged for model predictive control (MPC), which employs real-time optimization to a cost function within a finite horizon. Although MPC still presents similar limitations near the linear regions for some flows --such as difficulties in stabilizing the system--
it has been successfully applied to different flow control problems \citep{bieker2020deep, arbabi2018data, deda2023backpropagation, morton2018deep}. For example, near stabilization of the flow past a cluster of cylinders is achieved by \citet{maceda2021stabilization}.

The determination of optimal or advantageous locations to place sensors to enable full state reconstruction has been explored through a variety of methods, including via gappy proper orthogonal decomposition \citep{willcox2006unsteady}, a pivoted QR decomposition of an identified set of basis functions \citep{manohar2018data}, data-driven dynamical models with Kalman filters \citep{sashittal2021data,graff2023information}, and shallow decoder neural networks \citep{williams2022data}. As well as state reconstruction, the problem of sensor placement arises in flow control problems, where the placement of both sensors and actuators are important, and can be optimised and analysed using linear control theory \citep{chen2011h2,jin2022optimal, freire2020actuator}. 
The sensor placement methodology employed in the present work differs from these previous methods, and utilises a modified neural network loss function that incorporates the L1 norm of node (and thus physical location) weights. The use of L1-norm-based loss functions has been applied to find sparse solutions in a range of flow analysis problems, such as to find a reduced set of DMD modes to represent flow dynamics \citep{jovanovic2014sparsity}, and to find spatially-localised resolvent modes \citep{skene2022sparsifying, lopez2023sparsity}.

In the present work, we present a novel approach that enables nonlinear control design by utilizing neural network architectures to model both the system and control law. To achieve stabilization of complex systems exhibiting nonlinear dynamics, a neural network controller (NNC) is trained in closed loop with a neural network surrogate model (NNSM), which is built from data obtained from a real plant \citep{deda2023backpropagation}. The methodology is proposed as a candidate solution to overcome the aforementioned lack of measurements near the linear region of unstable nonlinear systems. The training framework also allows for optimal sensor placement by automatic selection from a set of candidate probes. Here, through penalization of model inputs, a sparse configuration is aimed that exclude the less relevant sensors. Furthermore, a byproduct consisting of datasets that contain measurements from approximately linear regions of the state space is leveraged for the data-driven estimation of equilibrium points---which are used as control setpoints---and for conducting a data-driven linear stability analysis. This novelty allows for approaching both problems without the need to modify the solver and enables the construction of linearised flow models involving complex geometries, as well as other nonlinear systems. Moreover, it potentially enables equilibrium estimation and stability analysis through data gathered from experimental applications.
The capabilities of the proposed methodology are demonstrated through analysis and control of four different plants, namely the Lorenz system, a modified version of the Kuramoto-Sivashinsky equation, a streamwise-periodic 2D channel flow, and a confined cylinder flow.

The remaining of this work is organized as follows: Section \ref{sec:methods} describes the tools and techniques introduced and employed along the work, while Section \ref{sec:training} describes in detail the proposed iterative approach for training the NNs. In Section \ref{sec:cases}, we describe the different systems/plants to which the techniques are applied. Section \ref{sec:results} then presents the results for each choice of plant. Finally, the benefits and limitations of the approaches presented are discussed in Section \ref{sec:conclusions}. Finally, the Appendix \ref{appA} describes the hyperparameters for building and training the neural networks for each case studied.


 
\section{Methodology} \label{sec:methods}
\subsection{Neural network surrogate models}

In this work, NNSMs are employed to represent the dynamics of complex systems, such as fluid flows. Given a fixed time step $\Delta t$, we are interested in discrete systems of the form
\begin{equation}
    \mathsfbi{x}_{k+1} = F(\mathsfbi{x}_k,\mathsfbi{u}_k) \mbox{ ,}\label{eq:fom}
\end{equation}
where $\mathsfbi{x}$ is the vector containing the system states, $\mathsfbi{u}$ is the vector of control inputs and the subscripts represent the discrete time steps such that time $t=k\Delta t$. The states vector $\mathsfbi{x}$ consists of a set of sensor measurements from which the dynamics can be inferred using data-driven methods. This set of of measurements consists of a subset of the true plant full set of states.

From a dataset containing measurements of both $\mathsfbi{x}$ and $\mathsfbi{u}$ from the plant, for multiple time steps, a NNSM is trained to approximate $F(\mathsfbi{x},\mathsfbi{u})$. A schematic of the neural network, with its inputs and outputs, is presented in figure \ref{fig:nn_rom}. Here, the symbol $\tilde{F}$ is used to denote the  function represented by the NNSM, and thus a good fit would imply $\tilde{F} \approx F$. The general approach is similar to that described in  \cite{deda2023backpropagation}.
%
The network is composed of fully connected hidden layers which employ a rectified linear unit (ReLU) activation function, whereas a linear function is used for the output layer.
The data for training and prediction are normalised so the average for each input and output is 0 and the standard deviation is 1. 
The loss function used to train this NNSM is discussed in section \ref{sec:sparse}. We also point out that the present NNs have a simple complexity, being kept at a maximum of two hidden layers for the cases studied in this work (with further details provided in Appendix \ref{appA}).

\subsection{Sensor placement}\label{sec:sparse}

Another idea explored and applied in the current work is related to the problem of sensor placement. 
There are several reasons to incorporate sensor placement into the proposed methodology. First, using a smaller number of sensors reduces the total number of training parameters in the NNSM, which in turn reduces the amount of training data required. Second, real-world flow control applications typically use limited real-time measurements, and selecting a small number of sensors that are optimal for control has practical advantages. Lastly, the incorporation of sensor placement into the NN training process allows for optimal sensor locations to be identified without relying on any prior knowledge or intuition of the system dynamics.

The approach proposed here consists of searching for a subset of relevant sensors among an initial set of candidates. To do so, we implement a layer of trainable parameters that are used to weight the NNSM inputs. These weights are penalised using an L1 regularization. The main objective is the production of a sparse layer excluding unnecessary measurements. Hence, after training, the weights below a given threshold (in absolute values) are truncated to 0. For the initialization, each weight value is set to 1.

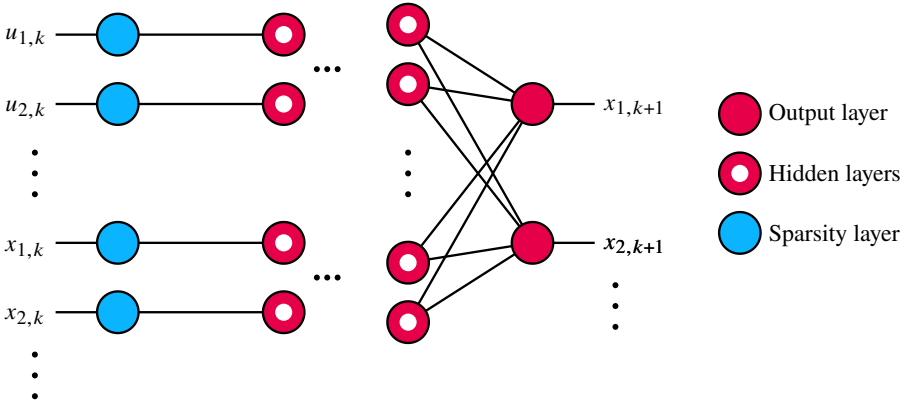
\begin{figure}
\centering
\begin{overpic}[width=.88\linewidth,trim={3.4cm 2cm 0cm 2cm},clip]{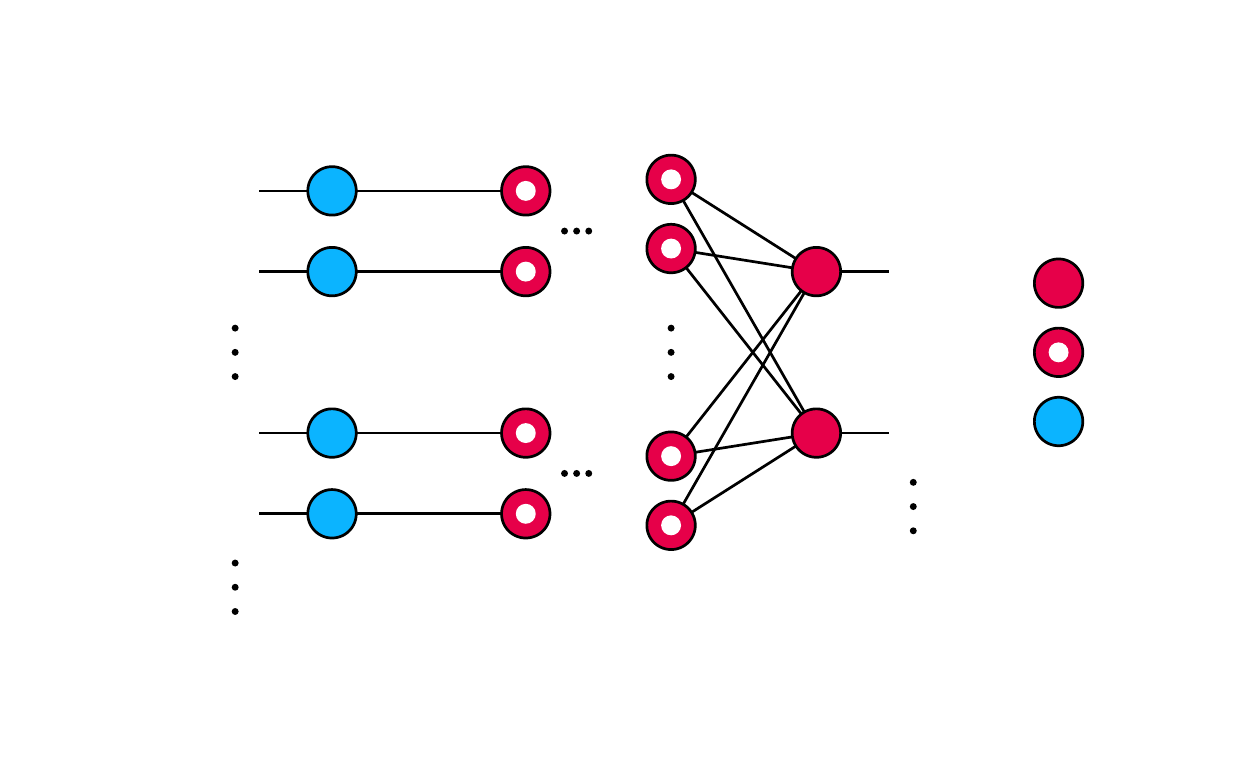}
\put(0,445){$u_{1,k}$}
\put(0,365){$u_{2,k}$}
\put(0,210){$x_{1,k}$}
\put(0,135){$x_{2,k}$}
\put(667,367){$x_{1,k+1}$}
\put(667,213){$x_{2,k+1}$}
\put(667,213){$x_{2,k+1}$}
\put(850,355){\footnotesize{Output layer}}
\put(850,288){\footnotesize{Hidden layers}}
\put(850,221){\footnotesize{Sparsity layer}}
\end{overpic}
\caption{Schematic of the NNSM where the red (blue) nodes are related to weights penalised by the L2 (L1) regularization. A ReLU function is employed in the nodes with a white mark, while a linear function is used for the other nodes.}
\label{fig:nn_rom}
\end{figure}

To avoid these weights getting too low by the cost of growing hidden layer weights, an L2 regularization is applied to the hidden layers. This ensures that the input weights can only decrease if they are irrelevant to estimate the output, which is still the full set of states in a future time instant. Figure \ref{fig:nn_rom} illustrates such configuration. The 
loss function 
\begin{equation}
    \mathcal{L} = \frac{1}{n}{\mathsfbi{g}\cdot \mathsfbi{g}} + r_2 {\mathsfbi{w}_\mathrm{h}\cdot \mathsfbi{w}_\mathrm{h}} + r_1 {\sum \lvert \mathsfbi{w}_\mathrm{s} \rvert}\mbox{ ,}
\end{equation}
is used for evaluation of the network convergence, where $\mathsfbi{g}$ is the array of differences between the labelled data and the NNSM output, $\mathsfbi{w}_\mathrm{h}$ is the stacked weights array for the hidden layers, $\mathsfbi{w}_\mathrm{s}$ is the array of weights for the sparsity layer, $n$ is the total number of candidate sensors, and $r_1$ and $r_2$ are the L1 and L2 regularization factors. In practice, $r_1$ can be adjusted such that larger values tend to provide sparser configurations at the cost of possibly reducing the accuracy of the model. The parameter $r_2$ can be increased to avoid the hidden layer weights to grow indefinitely larger to compensate for the sparsity layer weights becoming small.

In practice, the final model can be viewed as a function of a reduced vector $\mathsfbi{x_r}$ containing $n_r \leq n$ values, 
corresponding to the nonzero components of the sparsity weights, $\mathsfbi{w}_\mathrm{s}$.
Although this is not done in the current work, pruning could also be conducted to reduce the complexity of the neural network by excluding unnecessary connections \citep{zahn2022pruning}. Regardless of pruning, it is now possible to evaluate the NNSM as a function with the following form
\begin{equation}
    \mathsfbi{x}_{k+1} = \tilde{F}_r(\mathsfbi{x_r}_k,\mathsfbi{u}_k) \mbox{ ,}
\end{equation}
so the full state vector is reconstructed from the reduced set of measurements. 
Similarly, we can define a  function that predicts the components of the state corresponding to identified sensor locations,
\begin{equation}
    \mathsfbi{x_r}_{k+1} = \tilde{F}_{rr}(\mathsfbi{x_r}_k,\mathsfbi{u}_k) \mbox{ .}
\end{equation}
In short, $\tilde{F}_r$ and $\tilde{F}_{rr}$ are evaluated with the reduced set of states, but the former outputs the full set of states while the latter outputs the reduced set. Both $\tilde{F}_r$ and $\tilde{F}_{rr}$ are used for different goals in the present work, as it will be clear in the following sections.

\subsection{Estimation of equilibrium state} \label{sec:eq}
To tackle equilibrium estimation, a linearization of the NNSM is obtained through backpropagation. This linearization approach was first introduced by \citet{deda2023backpropagation} and, here, an improved technique is proposed. In the first step, the reduced states matrix $\mathsfbi{A_{rr}}_{(n_r \times n_r)}$ is obtained for the time invariant system of the form
\begin{align}\label{eq:linear}
    \mathsfbi{d_r}_{k+1} &= \mathsfbi{A_{rr}}\mathsfbi{d_r}_k + \mathsfbi{B_r}\mathsfbi{u}_k \mbox{ ,}\\
    \mathsfbi{d_r}_k &= \mathsfbi{x_r}_k - \mathsfbi{x_r}^*\mbox{ ,}\label{eq:diff}
\end{align}
where $\mathsfbi{A_{rr}}_{(n_r \times n_r)}$ and $\mathsfbi{B_r}_{(n_r\times m)}$ are constant matrices obtained from the Jacobian of $\tilde{F}_{rr}$ evaluated at a given equilibrium operating condition $\mathsfbi{x_r}^*$ and $\mathsfbi{u}^*$ as follows:
\begin{equation}
\nabla\tilde{F}_{rr}=
\begin{bmatrix}
\mathsfbi{A_{rr}} & \mathsfbi{B_r}
\end{bmatrix}
=
\begin{bmatrix}
\frac{\partial\tilde{F}_{rr}}{\partial \mathsfbi{x}} & \frac{\partial\tilde{F}_{rr}}{\partial \mathsfbi{u}}
\end{bmatrix}
=
\begin{bmatrix}
\frac{\partial\tilde{f}_1}{\partial{x}_1} &
\dots &
\frac{\partial\tilde{f}_1}{\partial{x}_{n_r}} & \frac{\partial\tilde{f}_1}{\partial{u}_1} &
\dots &  
\frac{\partial\tilde{f}_1}{\partial{u}_{m}}\\
\vdots & \ddots & \vdots & \vdots & \ddots & \vdots\\
\frac{\partial\tilde{f}_{n_r}}{\partial{x}_1} &
\dots &
\frac{\partial\tilde{f}_{n_r}}{\partial{x}_{n_r}} & \frac{\partial\tilde{f}_{n_r}}{\partial{u}_1} &
\dots &  
\frac{\partial\tilde{f}_{n_r}}{\partial{u}_{m}}
\end{bmatrix}\mbox{ .} \label{eq:jac}
\end{equation}
Here, $\tilde{f}_i$ is the $i$-th element in the output of $\tilde{F}$. 

In this work, $\mathsfbi{u}^* = 0$ is considered for calculation of a natural equilibrium point of the system, i.e., the equilibrium point related to the uncontrolled system. To compute $\nabla\tilde{F}_{rr}$, an initial guess $\mathsfbi{x_r}^*_0$ is used. The first estimate for the matrix $\mathsfbi{A_{rr}}$ is obtained by taking the first $n_r$ 
columns of $\nabla\tilde{F}_{rr}$, while the remaining columns compose $\mathsfbi{B_r}$ (see equation \ref{eq:jac}). The Newton method is employed to search for an equilibrium point for $\tilde{F}_{rr}$.

Since $\mathsfbi{u}^* = 0$, the system of the form
\begin{equation}
    \mathsfbi{x_r}_{k+1} = \mathsfbi{A_{rr}}\mathsfbi{x_r}_{k} + \mathsfbi{b_r}
\end{equation}
is considered, where $\mathsfbi{b_r}$ is an invariant vector. In this context, $\mathsfbi{b_r}$ is introduced to add a bias that shifts equilibrium from the origin. For a given equilibrium point $\mathsfbi{x_r}^*$,
\begin{equation}
    \mathsfbi{x_r}^* = \mathsfbi{A_{rr}}\mathsfbi{x_r}^* + \mathsfbi{b_r} \mbox{ ,}
\end{equation}
which gives
\begin{equation}
    \mathsfbi{x_r}^* = (I-\mathsfbi{A_{rr}})^{-1}\mathsfbi{b_r}\mbox{ .} \label{eq:update}
\end{equation}
The first guess for $\mathsfbi{b_r}_0$ is obtained by evaluating
\begin{equation}
\label{eq:br}
    \mathsfbi{b_r}_0 = \tilde{F}_{rr}(\mathsfbi{x_r}^*_0, \mathsfbi{0}) - \mathsfbi{A_{rr}}\mathsfbi{x_r}^*_0\mbox{ ,}
\end{equation}
which comes from the linear approximation of $\tilde{F}_{rr}$. With $\mathsfbi{b_r}$, $\mathsfbi{x_r}^*$ can be updated by evaluating equation \ref{eq:update}, which can subsequently be used to update $\mathsfbi{b_r}$ via equation \ref{eq:br}. The iterative process can be repeated to obtain an estimate of the equilibrium point. If desired, the equilibrium point can be mapped to the full set of sensors by evaluating
\begin{equation}
    \mathsfbi{x}^* = \tilde{F}_r(\mathsfbi{x_r}^*,\mathsfbi{u}_k) \mbox{ .}
\end{equation}
The complete procedure is synthesised in Algorithm \ref{alg:equilibrium_estimation}. 
Rather than iterating for a fixed number of steps, one could alternatively employ a stopping criteria based on the size of the update, or estimated error.
It is important to mention that this approach is better suitable when the spectrum of $\mathsfbi{A_{rr}}$ does not contain eigenvalues at 1+0j (integrator poles). In these cases, there is a continuous space of possible equilibrium points, any of which can be found by applying the Newton method.  
An alternative method for refining the equilibrium estimate, which is based on dynamic mode decomposition with control (DMDc) and which we observe can give improved results, will be described in Section \ref{sec:stability}. 

\begin{algorithm}
\caption{
Equilibrium estimation using linearization of the NNSM}
\label{alg:equilibrium_estimation}
\begin{algorithmic}[1]
    \State Obtain reduced states matrix $\mathsfbi{A_{rr}}$ and $\mathsfbi{B_r}$ using equation \ref{eq:jac}
    \State $\mathsfbi{b_r}_0 \gets \tilde{F}_{rr}(\mathsfbi{x_r}^*_0, \mathsfbi{0}) - \mathsfbi{A_{rr}}\mathsfbi{x_r}^*_0$ \Comment{Compute $\mathsfbi{b_r}_0$ using equation \ref{eq:br}}
    \State $\mathsfbi{x_r}^*_1 \gets (I - \mathsfbi{A_{rr}})^{-1}\mathsfbi{b_r}_0$ \Comment{Update $\mathsfbi{x_r}^*$ using equation \ref{eq:update}}
    \For{$i = 1$ to $N-1$} \Comment{Iterative process}
        \State $\mathsfbi{b_r}_i \gets \tilde{F}_{rr}(\mathsfbi{x_r}^*_{i}, \mathsfbi{0}) - \mathsfbi{A_{rr}}\mathsfbi{x_r}^*_{i}$ \Comment{Update $\mathsfbi{b_r}$ using equation \ref{eq:br}}
        \State $\mathsfbi{x_r}^*_{i+1} \gets (I - \mathsfbi{A_{rr}})^{-1}\mathsfbi{b_r}_i$ \Comment{Update $\mathsfbi{x_r}^*$ using equation \ref{eq:update}}
    \EndFor
    \State $\mathsfbi{x}^* \gets \tilde{F}_r(\mathsfbi{x_r}^*_{N}, \mathsfbi{u}_k)$ \Comment{Map to full set of sensors}
    \State \textbf{return} $\mathsfbi{x}^*$
\end{algorithmic}
\end{algorithm}

\subsection{Neural network controller}
\begin{figure}
\centering
\begin{overpic}[width=.9\linewidth]{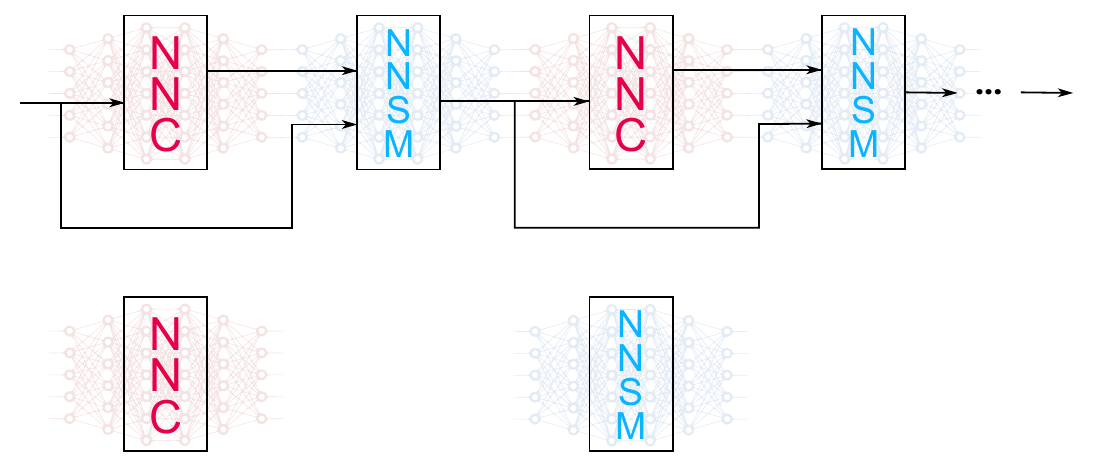}
\put(200,75){\footnotesize{= NN Controller $K$}}
\put(628,75){\footnotesize{= NN Surrogate Model $\tilde{F}_{rr}$}}
\put(36,345){\footnotesize{$\mathsfbi{{x}}_{\mathsfbi{r}_{0,j}}$}}
\put(446,345){\footnotesize{$\mathsfbi{\tilde{x}}_{\mathsfbi{r}_{1,j}}$}}
\put(840,360){\footnotesize{$\mathsfbi{\tilde{x}}_{\mathsfbi{r}_{2,j}}$}}
\put(930,360){\footnotesize{$\mathsfbi{\tilde{x}}_{\mathsfbi{r}_{n_h,j}}$}}
\put(230,375){\footnotesize{$\mathsfbi{u}_{0,j}$}}
\put(655,375){\footnotesize{$\mathsfbi{u}_{1,j}$}}

\end{overpic}
\caption{Schematic of the NNC training where the initial dataset is propagated through recurrent evaluations of the NNSM and NNC. Only the NNC weights and biases are updated during training in order to bring the states $\mathsfbi{x_r}_{1 \leq i \leq n_h,j}$  closer to $\mathsfbi{x_r}^*$.}
\label{fig:nnc_scheme}
\end{figure}
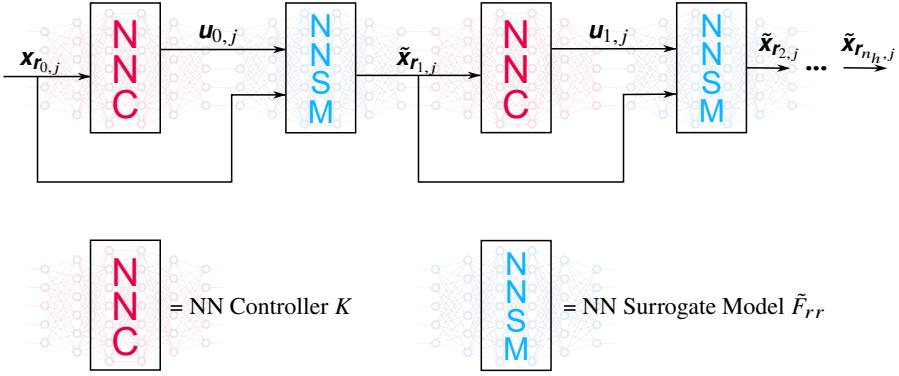

Neural network controllers (NNC) can be employed for the task of controlling complex nonlinear systems \citep{deda2023backpropagation}. A nonlinear control law of the form
\begin{equation}
    \mathsfbi{u}_k = K(\mathsfbi{x_r}_k) \label{eq:law}
\end{equation}
is implemented as a machine learning model, where the reduced set of states $\mathsfbi{x_r}_k$ are real-time measurements used to evaluate the control signal. A recurrent network is built for the training task as shown in figure \ref{fig:nnc_scheme}. The model is trained such that a set of initial conditions is brought closer to the equilibrium point $\mathsfbi{x_r}^*_0$ along iterations within a fixed discrete horizon $n_h$.

The training data consists of a set of measurements $\mathsfbi{X}_{0} = [\mathsfbi{x_r}_{0,1}\  \mathsfbi{x_r}_{0,2}\  \dots\  \mathsfbi{x_r}_{0,p}]$ containing $p$ samples of the $n_r$-state system at an initial time $k=0$. Note that the subscript 0 refers to the initial condition in the context of the finite-time closed-loop recurrent network, not being related to the simulation initial condition. The following loss function is targeted for minimization:
\begin{align}
    \mathcal{L} &= \frac{1}{p\,n_h}\left(\frac{\mathsfbi{w_e}}{n}\cdot\sum_{j=1}^{p}\sum_{i=1}^{n_h}{\mathsfbi{e}_{i,j}\circ \mathsfbi{e}_{i,j}} + \frac{\mathsfbi{w_u}}{m}\cdot\sum_{j=1}^{p}\sum_{i=0}^{n_h-1}{\mathsfbi{u}_{i,j}\circ \mathsfbi{u}_{i,j}}\label{eq:cost_nnc}\right)\mbox{ ,}\\
    \mathsfbi{e}_{i,j}&= \mathsfbi{\tilde{x}}_{i,j} - \mathsfbi{x}^*\mbox{ ,}
\end{align}
where $p$ is the size of the training dataset. Here, the weight vector $\mathsfbi{w_e}$ has all elements equal to $1/n_r$, which averages the result of the sum using an equal weighting of the states. Similarly, $\mathsfbi{w_u}$ has all elements equal to $w_u$, a scalar hyperparamenter used to penalise the control inputs.

The hidden NNC layers are implemented with ReLU activation, except for the output layer that has a sigmoid activation, which is normalised to limit the control effort within the desired range. Once the training process is finished, the NNC block is used independently of the training loop to evaluate the control law in real time through equation \ref{eq:law}. In the present work, all of the NNCs are implemented with a single hidden layer containing 8 nodes, which means they are computationally inexpensive as required for real-time applications. Appendix \ref{appA} further describes the hyperparameters employed.

\subsection{Linear stability analysis} \label{sec:stability}

Linear stability theory can be employed to elucidate mechanisms underlying the generation and amplification of flow instabilities. Here, we aim to employ the proposed neural network models to conduct linear stability analyses of unsteady flows, thereby illuminating physical mechanisms associated with unstable equilibrium points. In particular, with such an approach, it will be possible to compute the frequencies and growth rates of the most unstable modes and their associated eigenfunctions.

In an unsteady flow, a Reynolds decomposition allows the splitting of the flow states $\boldsymbol{x}(\boldsymbol{x_c},t)$ in a base flow $\boldsymbol{x}^*(\boldsymbol{x_c})$, here obtained at the equilibrium point, plus a time-dependent fluctuation component $\boldsymbol{x}'(\boldsymbol{x_c},t)$. In the present notation, $\boldsymbol{x_c}$ represents the spatial coordinates. If the fluctuations are sufficiently small, the equations can be linearised about the base flow and the Navier--Stokes equations can then be written in the following linear form 
	\begin{equation}
	\frac{\partial \boldsymbol{x}'}{\partial t} = \boldsymbol{\mathcal{A}} \boldsymbol{x}' \mbox{ ,}
	\label{eq:LNS_1}
	\end{equation}
where the matrix $\boldsymbol{\mathcal{A}} = \boldsymbol{\mathcal{A}}(\boldsymbol{x}^*)$ is a linear operator. For a system that is homogeneous in the streamwise ($x_c$) and spanwise ($z_c$) directions, the evolution of linear disturbances by equation \ref{eq:LNS_1} can be investigated by a direct analysis of the operator $\boldsymbol{\mathcal{A}}$ with the transformation
	\begin{equation}
 \label{eq:FT}
	\boldsymbol{x}'(x_c,y_c,z_c,t) = \int_{-\infty}^{\infty} \int_{-\infty}^{\infty} \int_{-\infty}^{\infty} \hat{\boldsymbol{x}}(\alpha,y_c,\beta,\omega) e^{\mathrm{i} (\alpha x_c + \beta z_c-\omega t)} \, d \alpha \, d \beta \, d \omega \mbox{ ,}
	\end{equation}
where both the streamwise and spanwise wavenumbers $\alpha, \beta \in \mathbb{R}$, and the frequency $\omega \in \mathbb{C}$. 

In discrete form, and under the transformation of equation \ref{eq:FT}, equation \ref{eq:LNS_1} 
can be written as
	\begin{equation}
	-\mathrm{i} \omega \hat{\boldsymbol{v}} = \mathsfbi{A} \hat{\boldsymbol{v}} \mbox{ .}
	\end{equation}
In this case, the linear operator becomes $\mathsfbi{A} = \mathsfbi{A}(\boldsymbol{x}^*,\alpha,\beta)$, and it can be analysed for each separate wavenumber pair $(\alpha,\beta)$ as an eigenvalue problem 
     \begin{equation}
	 \mathsfbi{V} \bm{\Lambda} = \mathsfbi{A} \mathsfbi{V} \mbox{ .}
	 \label{eq:LSA}
	 \end{equation}
In this equation, the columns $\hat{\boldsymbol{v}}_j$ of $\mathsfbi{V}$ are the eigenvectors of $\mathsfbi{A}$, and the eigenvalues 
 $\lambda_j = -\mathrm{i} \omega_j$ are the corresponding diagonal entries of $\bm{\Lambda}$. 
 Here the frequency and growth rate are the real and imaginary parts of $\omega_j$, respectively. 

There are several ways in which the identified NNSMs and NNCs can be leveraged to conduct stability analyses of the underlying dynamical systems. Most directly, we could utilise the matrix $\mathsfbi{A_{rr}}$ arising from the linearization of the NNSM obtained through backpropagation (equation \ref{eq:jac}). Rather than linearising the global nonlinear model, we can alternatively identify a linear model directly from data taken near the equilibrium point, which is the method used in the present work. This approach amounts to performing a linear regression inspired in DMDc \citep{proctor2016dynamic}. The idea comes from the fact that the controlled system can be slightly perturbed, producing a rich set of data near the equilibrium point, where the dynamics of the system tends to be approximately linear.

Given a dataset $\mathsfbi{X} = [\mathsfbi{d}_0 \dots \mathsfbi{d}_{p-1}]$ containing $p$ consecutive full state measurements, a dataset $\mathsfbi{U} = [\mathsfbi{u}_0 \dots \mathsfbi{u}_{p-1}]$ containing the control inputs history, and a dataset $\mathsfbi{X}^\prime = [\mathsfbi{d}_1 \dots \mathsfbi{d}_p]$ containing full state measurements with a unit shift, matrices $\mathsfbi{A}$ and $\mathsfbi{B}$ can be inferred as follows:
\begin{align}
    \mathsfbi{G} &= \mathsfbi{X}^\prime \mathbf{\Omega} ^\dagger \mbox{ ,}\\
    \mathsfbi{G} &\approx \begin{bmatrix}
        \mathsfbi{A} & \mathsfbi{B} \\
    \end{bmatrix} \mbox{ ,}\\
    \mathbf{\Omega} &= \begin{bmatrix}
        \mathsfbi{X} \\
        \mathsfbi{U}
    \end{bmatrix} \mbox{ .}
\end{align}
As seen, matrix $\mathbf{\Omega}$ is built from $\mathsfbi{X}$ and $\mathsfbi{U}$. By calculating its Moore-Penrose inverse $\mathbf{\Omega} ^\dagger$, it is possible to compute $\mathsfbi{G}$, from which approximations for $\mathsfbi{A}$ and $\mathsfbi{B}$ can be extracted. 

Note that $\mathsfbi{X}$ is composed of the deviation of the states from their estimated equilibrium values, as shown in equation \ref{eq:diff}. Admitting that the estimation of this equilibrium $\mathsfbi{x}^*$ is not exact, the proposed linear regression can be contaminated. To account for such imperfect estimation, a modification to this method is proposed, where a matrix $\mathsfbi{O} = [1 \dots 1]_{(1 \times p)}$ containing ones is employed. This allows for the modified procedure:
\begin{align}
    \mathsfbi{H} &= \mathsfbi{X}^\prime \mathbf{\Psi} ^\dagger \mbox{ ,}\\
    \mathsfbi{H} &\approx \begin{bmatrix}
        \mathsfbi{A} & \mathsfbi{B} & \mathsfbi{c}\\
    \end{bmatrix} \mbox{ ,}\\
    \mathbf{\Psi} &= \begin{bmatrix}
        \mathsfbi{X} \\
        \mathsfbi{U} \\
        \mathsfbi{O} 
    \end{bmatrix} \mbox{ .}
\end{align}
For systems with no integrator poles, this modification allows for the estimation of a system of the form
\begin{equation}
    \mathsfbi{d}_{k+1} = \mathsfbi{A}\mathsfbi{d}_k + \mathsfbi{B}\mathsfbi{u}_k + \mathsfbi{c} \mbox{ ,}
\end{equation}
where $\mathsfbi{c}$ is a constant vector that displaces the equilibrium point of the new states $\mathsfbi{d}$ from 0. Beyond providing a better estimation of $\mathsfbi{A}$ to conduct stability analysis, the addition of the constant $\mathsfbi{c}$  allows for the correction of the previous estimate of $\mathsfbi{x}^*$. Although this is not done in the current work, it could also be used, for example, to compensate control loop measurements to improve NNC results. Since the NNC is trained to control the NNSM and, furthermore, $\mathsfbi{x}^*$ is an equilibrium point for the NNSM, $\mathsfbi{c}$ could be used to adapt the NNC loop for the real plant.

To reduce computational costs, two different sets of matrices are used based on the sparse configuration obtained as described in section \ref{sec:sparse}:
\begin{itemize}
    \item The first is obtained by using the reduced states dataset $\mathsfbi{X_r} = [\mathsfbi{d_r}_0 \dots \mathsfbi{d_r}_{p-1}]$. Replacing $\mathsfbi{X}$ by $\mathsfbi{X_r}$, the algorithm produces a matrix $\mathsfbi{A_r}_{(n \times n_r)}$ which maps the reduced set of sensors to the full set at the next timestep, as well as $\mathsfbi{B}$ 
    and $\mathsfbi{c}$.
    \item The second set of matrices is obtained by also using $\mathsfbi{X_r}$ with the further substitution of $\mathsfbi{X}^\prime$ by a reduced matrix $\mathsfbi{X_r}^\prime$. This produces the same matrices $\mathsfbi{A_{rr}}_{(n_r \times n_r)}$ and $\mathsfbi{B_r}_{(n_r\times m)}$ shown in section \ref{sec:eq}---with potentially better accuracy---as well as the reduced bias vector $\mathsfbi{c_r}$.
\end{itemize}

A stability analysis can be conducted by finding the eigenvectors $\mathsfbi{V_r}_i$ and eigenvalues $\lambda_i$ from the square matrix $\mathsfbi{A_{rr}}$, where $i=1,2,\dots,n_r$. Since each vector $\mathsfbi{V_r}_i$ is an eigenvector of $\mathsfbi{A_{rr}}$, each $\mathsfbi{A_{rr}}\mathsfbi{V_r}_i$ is also an eigenvector of $\mathsfbi{A_{rr}}$, which allows the mapping
\begin{equation}
    \mathsfbi{V}_{i(n\times 1)} = \mathsfbi{A_r}_{(n\times n_r)}\mathsfbi{V_r}_{i(n_r\times 1)}
\end{equation}
to estimate the eigenvectors across all sensor states. This mapping is employed in this work for visualization of modes projected over the full set of candidate sensors. 

\section{Iterative training} \label{sec:training}
A key aspect of the proposed methodology is the development of a training methodology that ensures that both the NNSM and NNC are trained with sufficient quantities of data, particularly in the regions of state space where model accuracy is important for control effectiveness.
To train and improve the neural network models employed in this work, an iterative training algorithm is proposed. Four different modes of data sampling are considered for training the NNSM and the NNC. Each mode produces data that are compiled for training both networks. They are:
\begin{itemize}
    \item \textit{Uncontrolled sweep} mode: the system is perturbed by an open-loop control signal with no closed-loop control;
    \item \textit{Controlled sweep} mode: the system is controlled in closed loop at the same time that it is perturbed by an open-loop signal;
    \item \textit{Release} mode: the plant runs with no control and no open-loop perturbation, allowing for better visualization of the next mode and to test whether the system can be successfully controlled from an initially uncontrolled state;
    \item \textit{Control} mode:  the system is controlled in closed loop with no open-loop perturbation.
\end{itemize}
All open-loop perturbations applied in this work are random staircase signals with constant step length.

The \textit{uncontrolled sweep} is the first mode employed so as to produce data with control input history. A number of samples is gathered from the perturbed plant. These first samples are used to compute the means and standard deviations used for data normalization of the neural networks inputs and outputs. This data is also used for training the NNSM, estimating its equilibrium point $\mathsfbi{x_r}^*$ and training the NNC, all for the first time. The NNSM sparsity layer has its negligible weights truncated according to a threshold (further described in Appendix \ref{appA}), which accounts for identifying and updating the relevant sensor locations.

The second mode to be run is the \textit{release} mode, which allows the system to achieve its natural (uncontrolled) behaviour, i.e., a limit cycle or a chaotic attractor, for example. 
The \textit{control} mode is applied after the system completely or partially returns to its natural operation. When the NNC is turned on, the dynamics of the controlled system drives the plant states to work under a new behaviour, which can tend to a new limit cycle, a new chaotic attractor or even to an equilibrium point - which can either be or not be desirably close to the estimated equilibrium, $\mathsfbi{x}^*$.

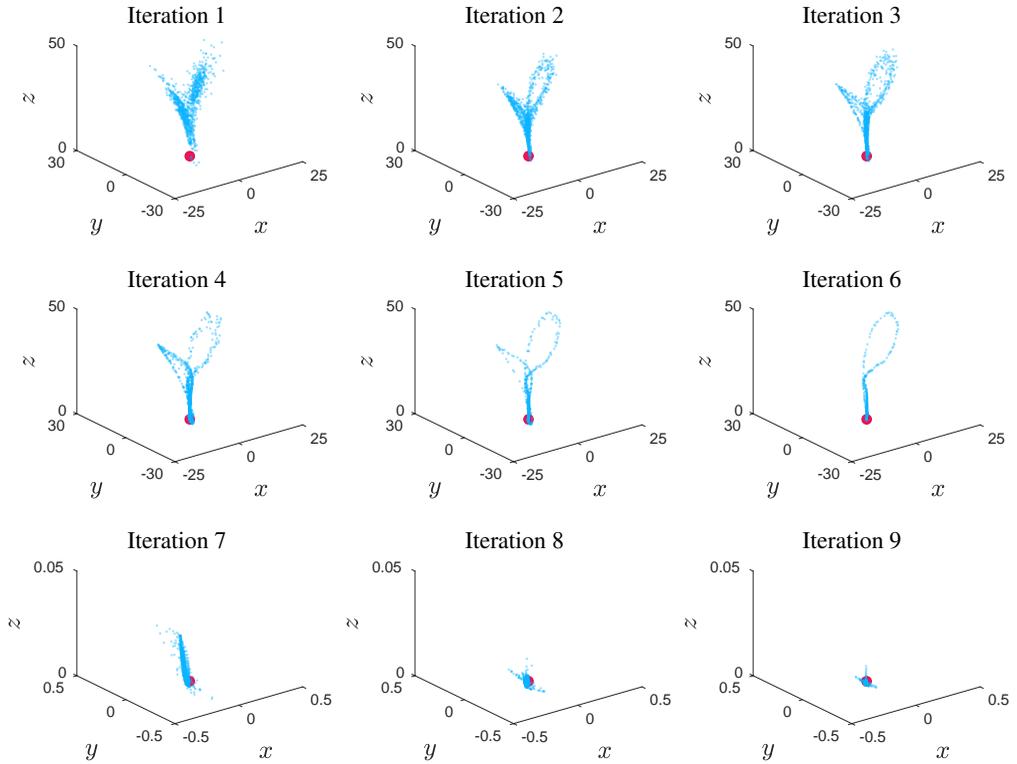
\begin{figure}
\centering
\begin{overpic}[width=.99\linewidth,trim={2cm 1cm 2cm 0.3cm},clip]{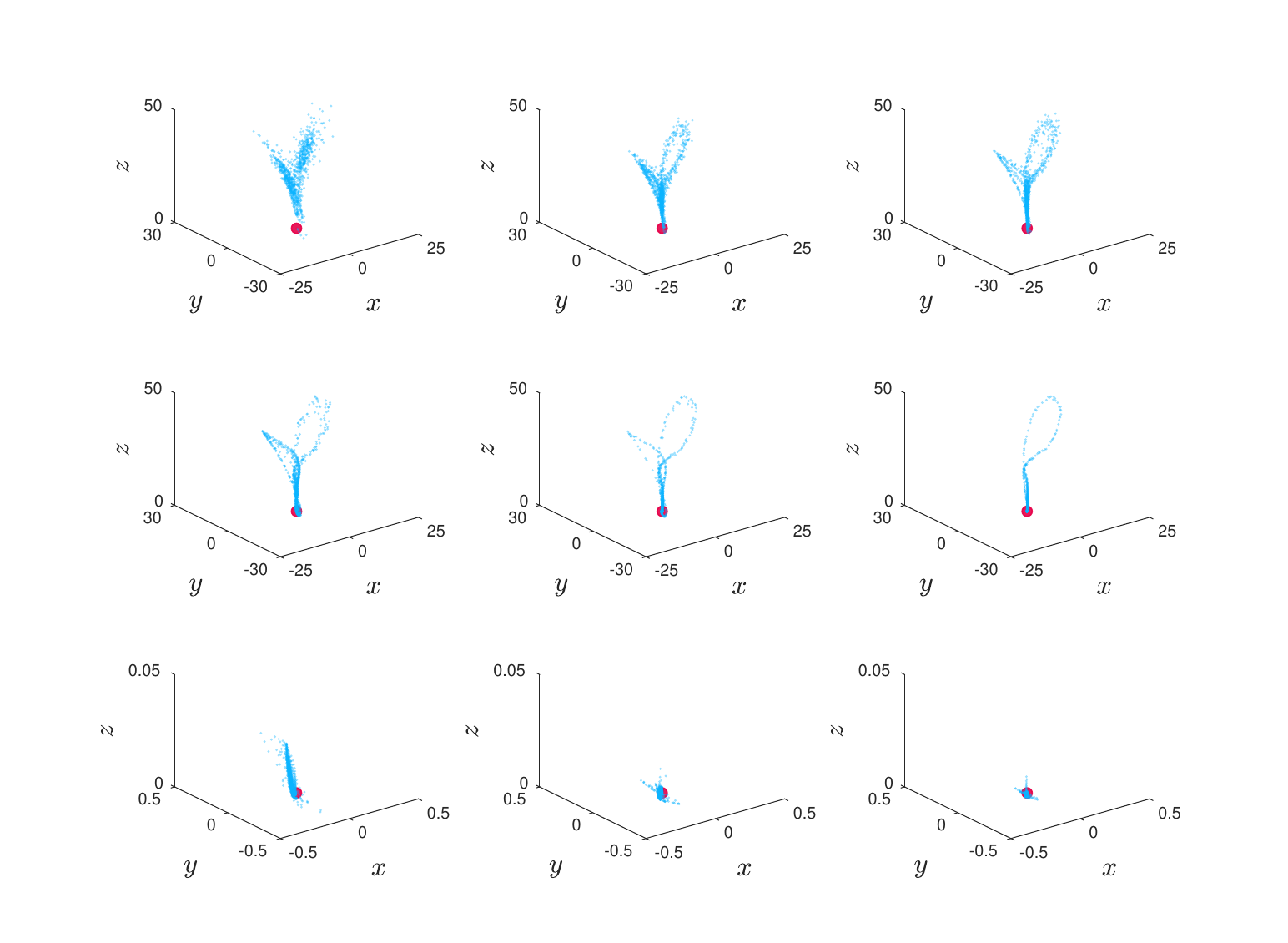}
\put(120,740){\small{Iteration 1}}
\put(454,740){\small{Iteration 2}}
\put(788,740){\small{Iteration 3}}
\put(120,480){\small{Iteration 4}}
\put(454,480){\small{Iteration 5}}
\put(788,480){\small{Iteration 6}}
\put(120,223){\small{Iteration 7}}
\put(454,223){\small{Iteration 8}}
\put(788,223){\small{Iteration 9}}
\end{overpic}
\caption{Phase portraits of the Lorenz system for data sampled during the \textit{sweep} modes at different iterations. Note that for iterations 7 to 9 the axes limits are considerably smaller. The red dot indicates the origin, which is an equilibrium point for the Lorenz system.}
\label{fig:iterative}
\end{figure}

Data is gathered in each of these three steps, building a database that grows along the iterative process. After they are run, the process is repeated for a number of iterations, but instead of doing an \textit{uncontrolled sweep}, the \textit{controlled sweep} mode is applied in every iteration beyond the first one. The idea is to leverage the fact that the controlled system should have a tendency to get closer to the equilibrium point, and by perturbing the controlled system appropriately, a larger set of data 
is obtained around this equilibrium. When a new iteration begins, the amplitudes of the open-loop perturbations are reduced by a constant factor $0<\alpha\leq1$, so that data is gathered closer to the actual plant equilibrium. At each iteration, the NNSM and the NNC are retrained with new data collected from the full database and the equilibrium point is updated using the retrained NNSM. Algorithm \ref{alg:data_sampling} synthesises the steps described, and a schematic showing the phase portrait for the Lorenz system is also presented in figure \ref{fig:iterative}. Each iteration in this figure depicts 1000 measurements sampled during the \textit{sweep} modes (\textit{uncontrolled} for iteration 1, \textit{controlled} for subsequent iterations). As the iterations advance, the closed-loop control tends to bring states closer to the origin, which is an equilibrium point for the Lorenz system. This enhances the estimate of the equilibrium point, and it can be seen in the smaller limits of the axes, as shown for the higher iterations. Also, beginning in iteration 7, since the perturbations are reduced between iterations, they become unable to shoot the controlled states out of the main stable orbit around the now-stabilised equilibrium. 
\begin{algorithm}[H] 
\caption{Data Sampling Algorithm}
\label{alg:data_sampling}
\begin{algorithmic}[1]
\For {$i=1$ to $n$}
\If {i==1}
\State Run \textit{uncontrolled sweep} mode to gather data
\State Compute parameters for normalization
\Else 
\State Reduce the amplitude of open-loop perturbations by a factor $\alpha$
\State Run \textit{controlled sweep} mode to gather data around the estimated equilibrium
\EndIf
\State Update the database with the new data
\State Train NNSM with the updated database
\State Truncate negligible weights in the sparsity layer
\State Estimate equilibrium from NNSM
\State Train NNC with the updated database\State Run release mode to allow the system to reach its natural behaviour
\State Gather data in \textit{release} mode
\State Run \textit{control} mode to drive the system to a new behaviour and test control effectiveness
\State Gather data in control mode
\EndFor
\end{algorithmic}
\end{algorithm}

For the stability analysis, only the data produced by the last \textit{sweep} mode is used since it may contain more accurate data near the equilibrium point, hopefully ruled by approximately linear dynamics. One of the advantages of using such type of data is enabling DMD-based techniques to identify structures corresponding to eigenvectors of the equillibirum-linearised system, even for
systems exhibiting strong nonlinearities.

\section{Study cases} \label{sec:cases}
This section outlines the four systems that will be used to test and validate the proposed methodology.
\subsection{Lorenz equations}

The first study case for this work is the Lorenz system, whose dynamics are described by the following set of nonlinear ordinary differential equations
\begin{align}
    \dot{x} &= \sigma(y-x)\mbox{ ,}\\
    \dot{y} &= \rho x -y -xz+u\mbox{ ,}\\
    \dot{z} &= -\beta z + xy\mbox{ ,}
\end{align}
where $\sigma = 10$, $\beta = 8/3$ and $\rho$ = 28 are chosen for the present analysis. With this set of parameters, the uncontrolled system ($u(t)$=0) behaves as a chaotic attractor. The goal is to apply the techniques proposed in this work to find an equilibrium point; to stabilise the system, bringing the states towards this point; and to compute a linear model to obtain the system poles. In this problem we do not conduct sensor placement.

\subsection{Modified Kuramoto-Sivashinsky equation}

The Kuramoto-Sivashinsky (KS) equation is given by the following nonlinear partial differential equation
\begin{equation}
    \frac{\partial v}{\partial t} + v\frac{\partial v}{\partial x_c} = -\frac{1}{R}\left( P\frac{\partial ^2 v}{\partial x_c^2} +\frac{\partial ^4v}{\partial x_c^4}\right) \mbox{ ,}
\end{equation}
where $R$ is equivalent to the Reynolds number in a fluid flow, and $P$ represents a balance between energy production and dissipation. This system has been used to emulate, for example, combustion instabilities in flame fronts \citep{sivashinsky1977derivation, kuramoto1978reaction} and hydrodynamic instabilities in shear flows \citep{fabbiane2014adaptive}. In this context, $x_c$ is the spatial coordinate and $v(x_c)$ is the variable of interest. 

It is clear that, for any constant $c$, $v(x_c) = c$ corresponds to an equilibrium point, i.e., $\partial v/ \partial x_c = 0$. Since we are interested---among other goals---in computing an equilibrium point for this system, we propose a modified version of the KS equation given by
\begin{equation}
    \frac{\partial v}{\partial t} + v\frac{\partial v}{\partial x_c} = -\frac{1}{R}\left( P\frac{\partial ^2 v}{\partial x_c^2} +\frac{\partial ^4v}{\partial x_c^4}\right) -\frac{Q}{L} \int_{0}^{L}(v(x_c)-V)\,dx_c + \sum_{i=0}^m B_i(x_c) u_i\mbox{ .} \label{eq:ks_mod}
\end{equation}
The new term $-Q/L \int_{0}^{L}(v(x_c)-V)\,dx_c$ corresponds to a spatial average of the difference between $v(x_c)$ and a chosen equilibrium $V$. We make use of this modification to ensure that the equilibrium point $v(x_c) = V$ is unique for $u_i(t)=0$. A globally unstable configuration is set by choosing $R=0.25$, $P=0.05$, $Q=0.0005$, $V=0.2$ and $L=60$ with periodic boundary conditions. The control term $\sum_{i=0}^m B_i(x_c) u_i$ comes similarly to the approach proposed by \citet{fabbiane2014adaptive}, where $m$ is the number of control inputs, $B_i(x_c)$ is a window function for each actuator and $u_i$ represents the control signals.

\begin{figure}
\centering
\includegraphics[width=.88\linewidth,trim={0.5cm 0cm 1.0cm 0cm},clip]{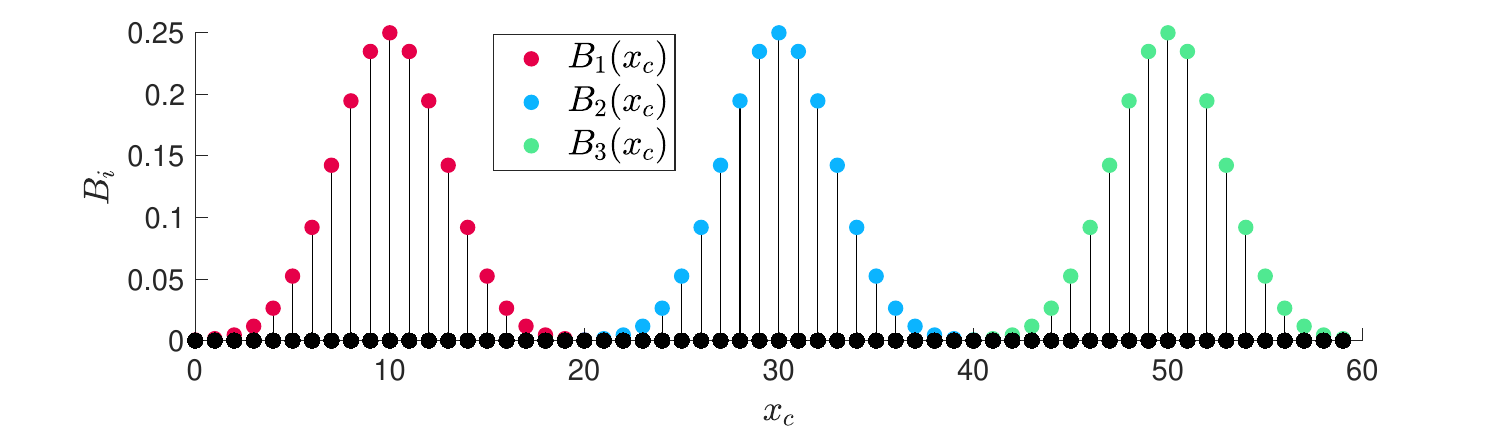}
\caption{Actuation and sensor setup for controlling the modified KS equation. The coloured dots represent the control input profiles $B_i$ at each of the 60 positions measured initially (candidate sensors).}
\label{fig:ks_setup}
\end{figure}

The equation is discretised in the present simulation with $\Delta x_c = 1$ and $\Delta t = 0.025$. Sampling for control is made each 400 timesteps, thus giving a control timestep $\Delta t_c = 10$. Three actuators are employed by using Gaussian window functions centered at $x_c = 10$, $x_c = 30$ and $x_c = 50$. Sampling of $v$ is conducted at each of the 60 discrete grid points. Figure \ref{fig:ks_setup} shows a sensor/actuation schematic. In this test case, we conduct equilibrium estimation, sensor placement, control and stability analysis.

\subsection{Periodic 2D channel flow}

\begin{figure}
\centering
\begin{overpic}[width=.9\linewidth,trim={6.8cm 5.5cm 5cm 5cm},clip]{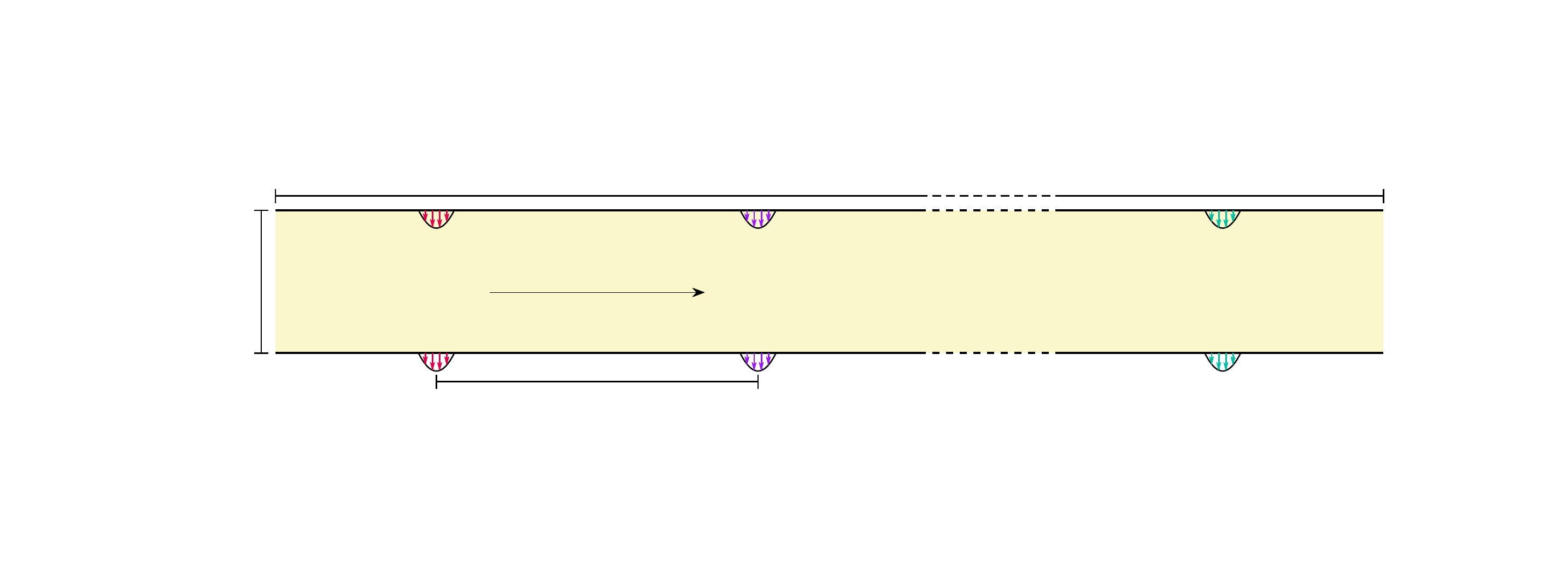}
\put(238,105){\scriptsize{Flow direction}}
\put(290,27){\scriptsize{$L/8$}}
\put(470,186){\scriptsize{$L=36h$}}
\put(8,92){\rotatebox{90}{\scriptsize{$2h$}}}
\put(171,95){\scriptsize{\color[HTML]{e60049}{$u_1$}}}
\put(441,95){\scriptsize{\color[HTML]{9b19f5}{$u_2$}}}
\put(831,95){\scriptsize{\color[HTML]{00bfa0}{$u_8$}}}
\put(52,68){\rotatebox{90}{\tiny{Periodic BC}}}
\put(965,68){\rotatebox{90}{\tiny{Periodic BC}}}
\end{overpic}
\caption{Actuation setup for controlling the 2D channel flow. Each pair of actuators works in opposition, ensuring zero-net mass flux.}
\label{fig:channel_setup}
\end{figure}
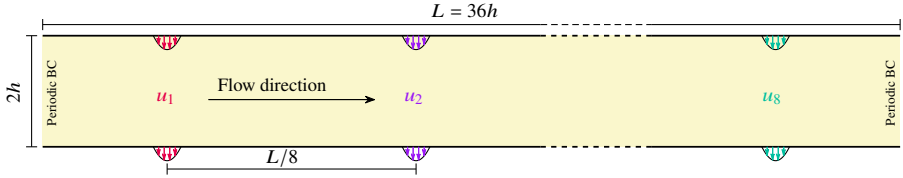

Another problem we consider is streamwise-periodic channel flow. We employ numerical simulations using an open source Nek5000 CFD code 
previously applied for RL control \citep{li2022reinforcement}
to solve the incompressible Navier-Stokes equations for a Reynolds number Re = 8000, 
based on the half channel height $h$ and the velocity at the channel center line $U$. The channel length is set to $L=36h$ so that two unstable modes appear, at streamwise wavelengths L/5 and L/6. A total of $m=8$ control inputs $u_1, \dots, u_8$ modulate the velocities of 8 pairs of actuators. Each pair works in opposition to ensure a zero-net mass flux, being weighted by a parabolic window function as depicted in figure \ref{fig:channel_setup}.

\begin{figure}
\centering
\includegraphics[width=.99\linewidth,trim={0.5cm 0cm 1.0cm 0cm},clip]{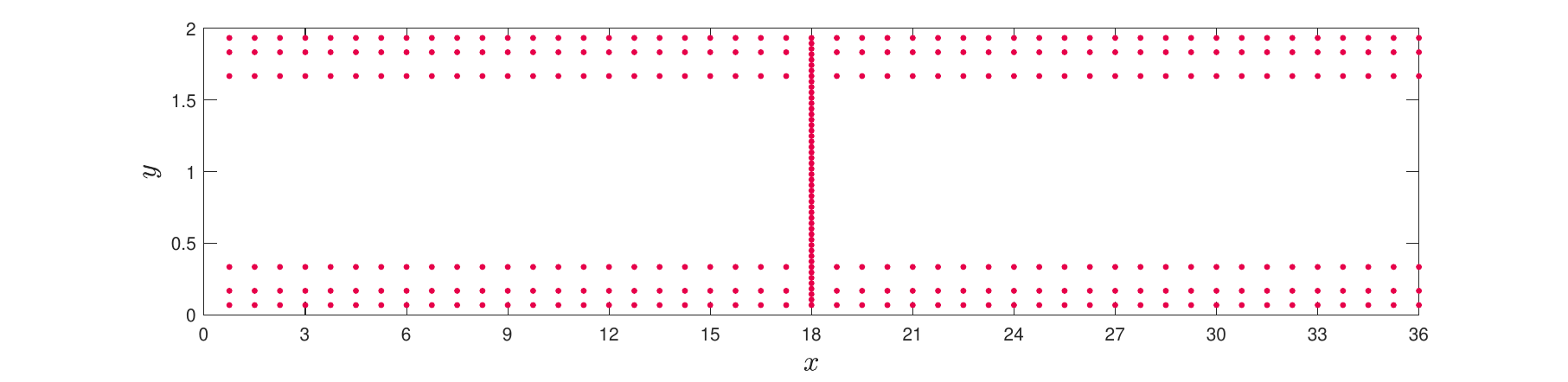}
\caption{Initial candidate sensor positions for the channel flow. The dimensions are distorted for better visualization.}
\label{fig:chanel_sensors}
\end{figure}

The distribution of candidate sensors is presented in figure \ref{fig:chanel_sensors}. 
The chosen locations are concentrated near the wall to capture velocity fluctuations from hydrodynamic instabilities. 
A vertical line of 50 sensors is also employed for comparison with stability analysis computed from the solution of the Orr-Sommerfeld equation. In total, 332 candidate sensor positions are chosen, where horizontal and vertical velocities $u(x_c,y_c)$ and $v(x_c,y_c)$ are sampled, totalling 664 signals. For this channel problem, we are interested in equilibrium estimation, sensor placement, control, and stability analysis. 

\subsection{Confined cylinder flow}

\begin{figure}
\centering
\includegraphics[width=.99\linewidth,trim={3cm 1.6cm 2cm 1cm},clip]{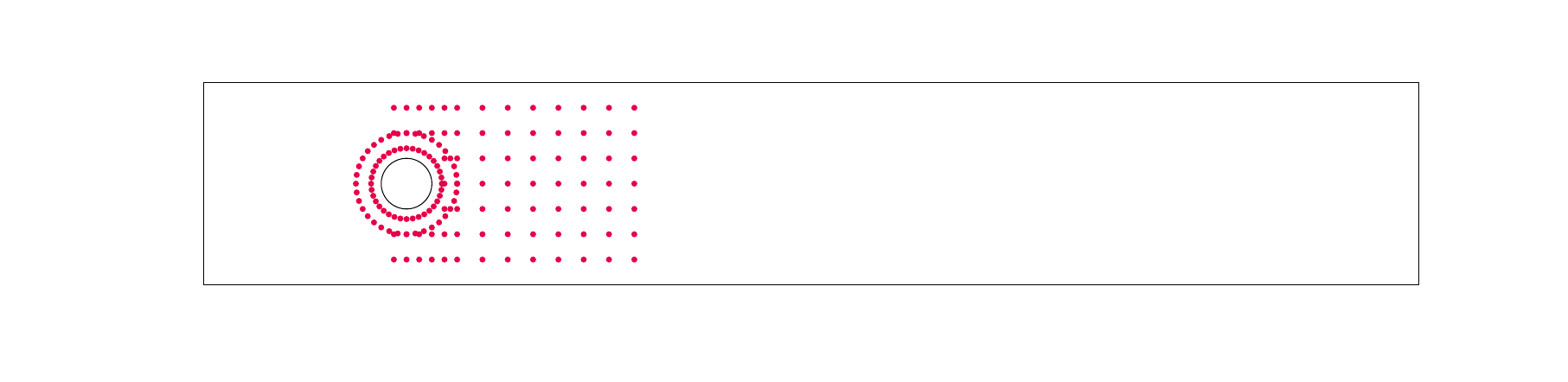}
\caption{Initial candidate sensor positions for the confined cylinder flow. The box represents the spatial domain used in the simulation.}
\label{fig:cyl_sensors}
\end{figure}

Finally, the study of a 2D confined flow past a cylinder is presented. The initial configuration of sensors and actuators is the same as that proposed by \citet{rabault2019artificial}. For this task, a Nek5000 setup \citep{li2022reinforcement} 
is employed considering a Reynolds number $\mathrm{Re}=150$ based on the cylinder diameter $D$ and maximum velocity $U$ of the inflow profile.
The initial configuration of sensors is equivalent to the one presented by \citet{rabault2019artificial}, depicted in figure \ref{fig:cyl_sensors}, where 153 locations are chosen, totalling 306 velocity measurements for $u$ and $v$. No-slip wall boundary conditions are applied at the top and bottom limits of the domain.

\begin{figure}
\centering
\begin{overpic}[width=.43\linewidth,trim={0cm 0cm 10cm 0cm},clip]{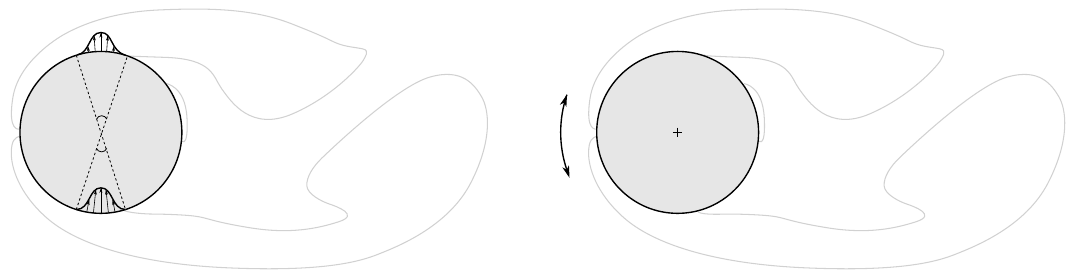}
\put(180,422){\footnotesize{$v_n$}}
\put(180,94){\footnotesize{$v_n$}}
\put(230,285){\tiny{$10\degree$}}
\end{overpic}
\caption{Actuation scheme applied to the cylinder flow. Blowing/suction devices in opposition are modulated by a single control input.}
\label{fig:cyl_act}
\end{figure}

The actuation scheme is presented in figure \ref{fig:cyl_act}. A single control input modulates minijets in opposition, thus providing zero-net mass flux. For this problem, we are interested in finding an equilibrium point (only to estimate a control setpoint), sensor placement and flow control.
 
\section{Results}\label{sec:results}

\subsection{Lorenz equations}

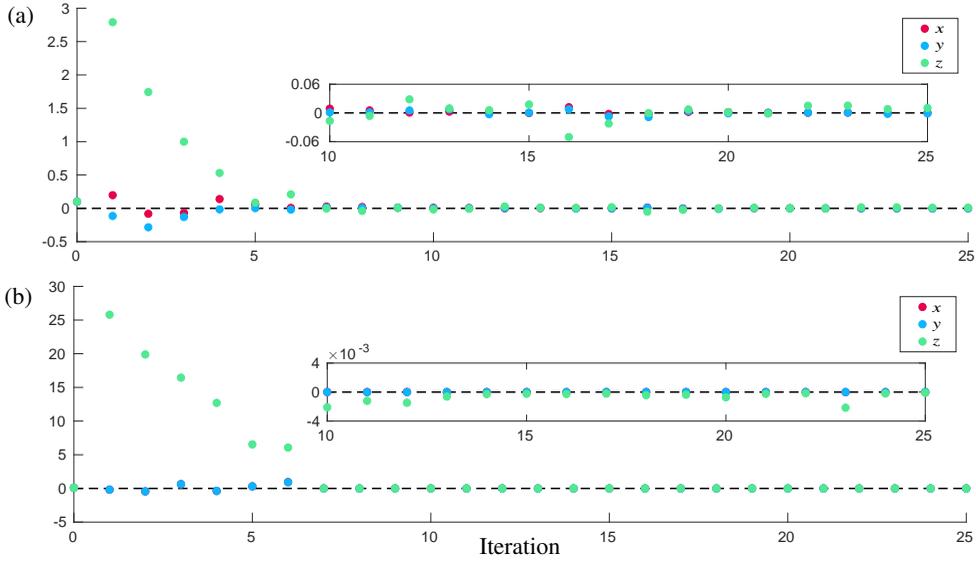
\begin{figure}
\centering
\begin{overpic}[width=.99\linewidth,trim={2cm 0cm 2cm 0.3cm},clip]{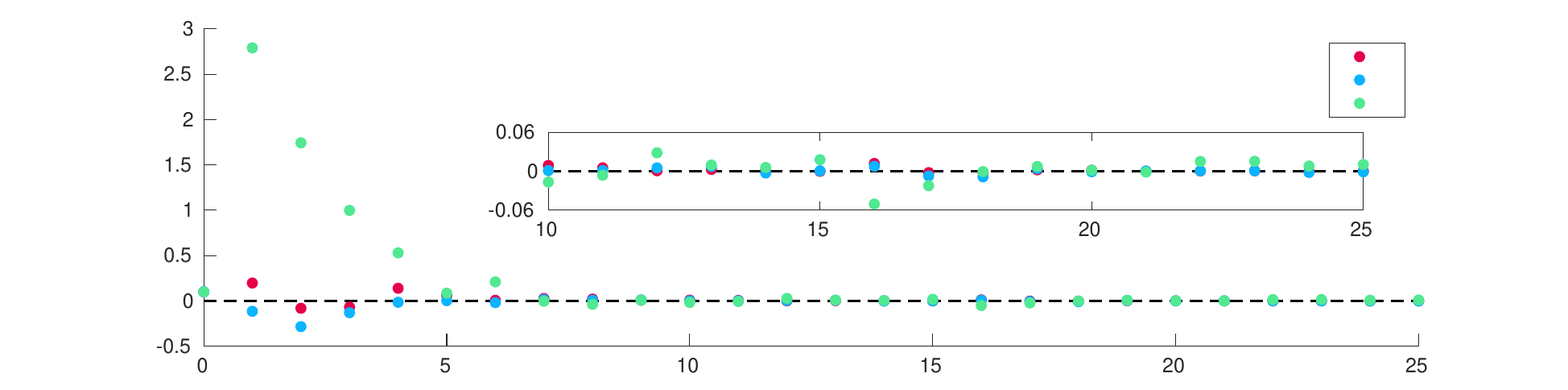}
\put(928, 237){\tiny{$x$}}
\put(928, 221){\tiny{$y$}}
\put(928, 204){\tiny{$z$}}
\put(10, 247){(a)}
\end{overpic}
\begin{overpic}[width=.99\linewidth,trim={2cm 0.0cm 2cm 0.3cm},clip]{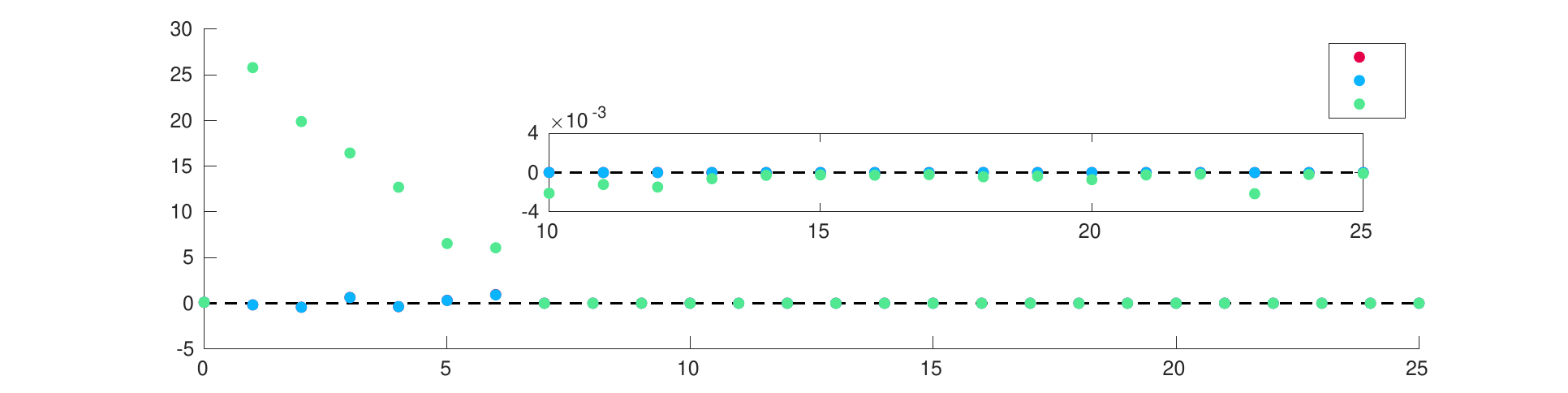}
\put(928, 239){\tiny{$x$}}
\put(928, 223){\tiny{$y$}}
\put(928, 206){\tiny{$z$}}
\put(480, 0){\footnotesize{Iteration}}
\put(10, 247){(b)}
\end{overpic}
\caption{Equilibrium point estimation along iterations. The estimation through backpropagation is shown in (a), where more accurate estimates are made at first iterations. The fixed estimation through the DMDc variant is shown in (b), where better approximations are found in the last iterations. In this case, the red dots are hidden behind the blue ones.}
\label{fig:lorenz_eq}
\end{figure}

To assess the performance 
of the proposed techniques, results for the low order Lorenz system are presented. Starting with the problem of estimating an equilibrium point for this system, figure \ref{fig:lorenz_eq} presents the values of each state at each iteration. Two plots compare this computation using both the backpropagation-based linearization of the neural network (as described in section \ref{sec:eq}) and the alternative version  employing the modified DMDc method proposed in section \ref{sec:stability}.
In both cases, a convergence towards equilibrium at $x=y=z=0$ is seen. It is clear that, for the initial iterations, better approximations are obtained using backpropagation (figure \ref{fig:lorenz_eq}(a)). The fact that DMD based techniques rely on data sampled from approximately linear systems compromise their performance when the plant is strongly nonlinear. As discussed by \citet{deda2023backpropagation}, the linearization of a nonlinear model at the point of interest can provide better results when nonlinearities cannot be neglected. On the other hand, for the later iterations, the DMD based approach is able to provide more accurate results, since the closed-loop control is able to maintain states near equilibrium, i.e., operating almost as a linear system (figure \ref{fig:lorenz_eq}(b)). 

\begin{figure}
\centering
\begin{overpic}[width=.99\linewidth,trim={2cm 0cm 2cm 0.3cm},clip]{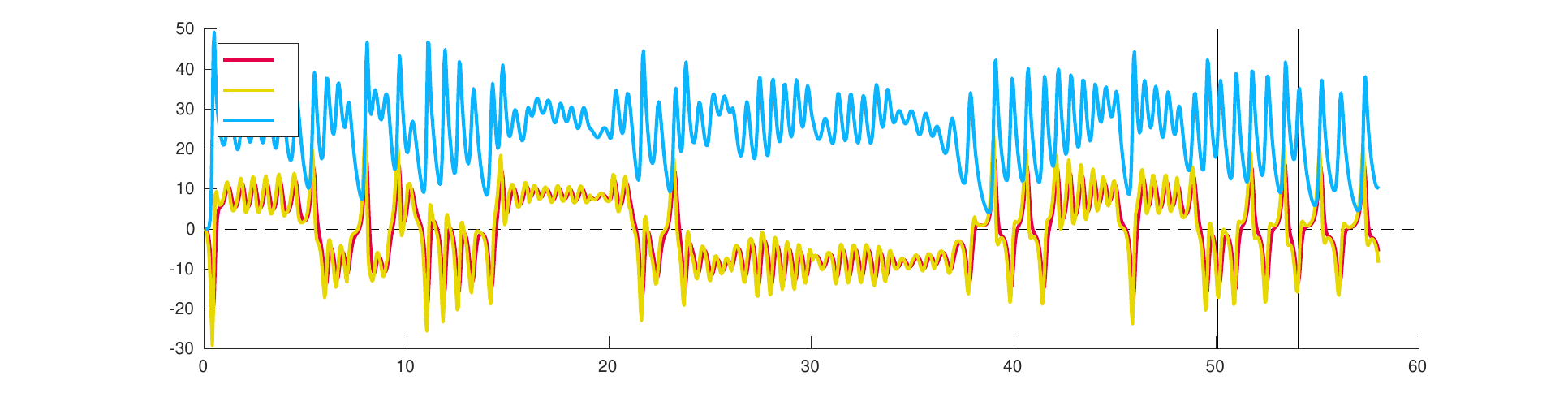}
\put(132, 238){\tiny{$x$}}
\put(132, 217){\tiny{$y$}}
\put(132, 196){\tiny{$z$}}
\put(450,255){\small{\textit{s.m.}}}
\put(825,255){\small{\textit{r.m.}}}
\put(893,255){\small{\textit{c.m.}}}
\put(10, 247){(a)}
\end{overpic}
\begin{overpic}[width=.99\linewidth,trim={2cm 0cm 2cm 0.3cm},clip]{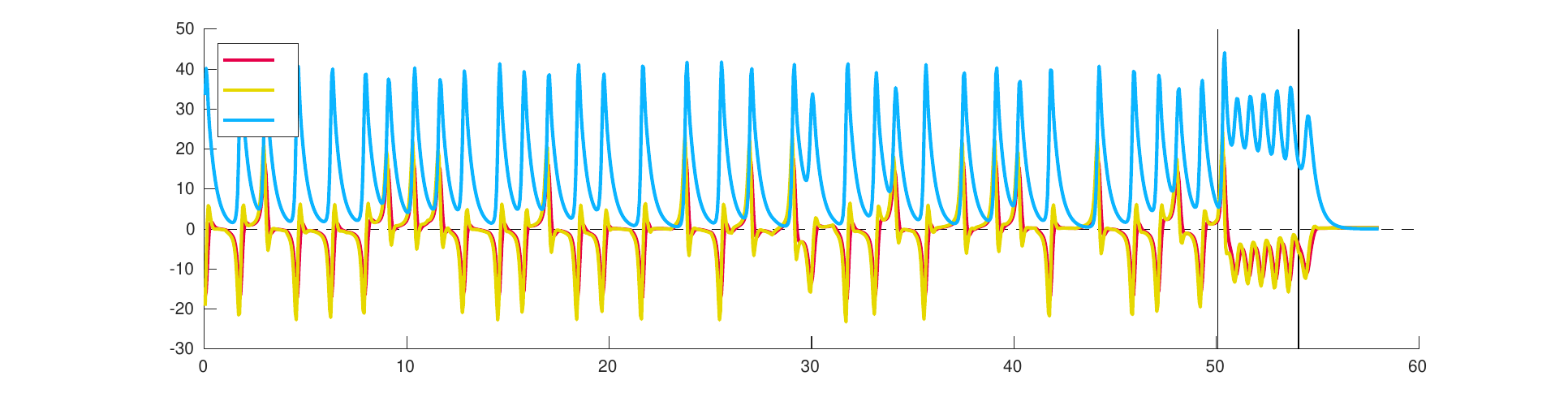}
\put(132, 239){\tiny{$x$}}
\put(132, 218){\tiny{$y$}}
\put(132, 197){\tiny{$z$}}
\put(450,256){\small{\textit{s.m.}}}
\put(825,256){\small{\textit{r.m.}}}
\put(893,256){\small{\textit{c.m.}}}
\put(420, 0){\footnotesize{Dimensionless time}}
\put(10, 260){(b)}
\end{overpic}
\caption{\textit{Sweep}, \textit{release} and \textit{control} modes (\textit{s.m.}, \textit{r.m.} and \textit{c.m.}, respectively) for iterations (a) 1 and (b) 4. The trained NNC is unable to stabilise the system in iteration 1, in contrast to the trained control system in iteration 4, where stabilization is achieved. This can be seen by comparing both control modes (\textit{c.m.}).}
\label{fig:lorenz_states}
\end{figure}
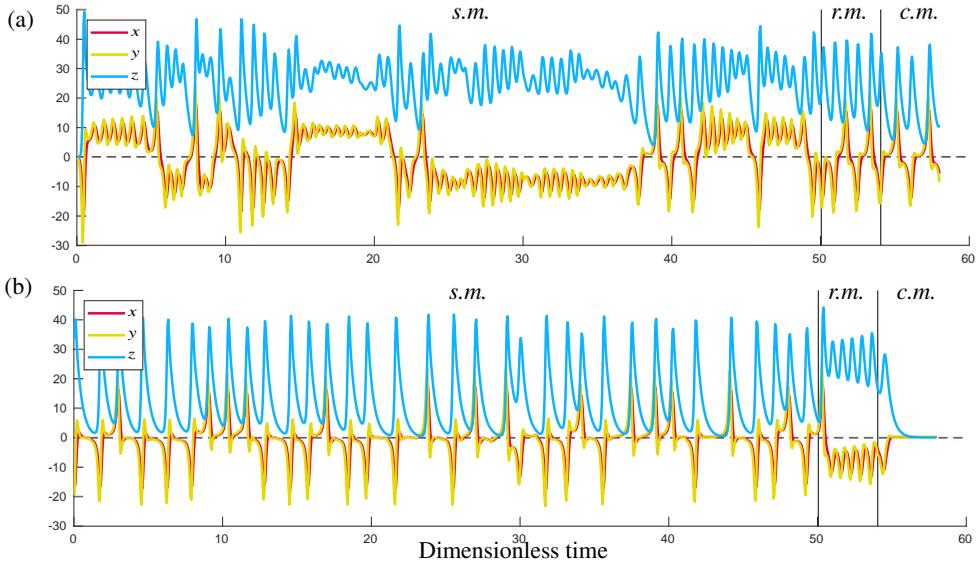

The stabilization of the controlled Lorenz system is depicted in figure \ref{fig:lorenz_states}, where signals are presented against time for iterations 1 and 4. For the former case, depicted in figure \ref{fig:lorenz_states}(a), it is still not possible to obtain a stabilizing controller. This might be due to poor estimation of the dynamics near the equilibrium point obtained for such iteration. On the other hand, at the iteration 4 shown in figure \ref{fig:lorenz_states}(b), the trained NNC is able to fully stabilise the plant, bringing the states close to $x=y=z=0$. As seen in figure \ref{fig:lorenz_eq}(a), the equilibrium point for iteration 4 is estimated at $x \approx 0.14$, $y \approx -0.01$ and $z \approx 0.53$, which is close enough to allow for the NNC to stabilise the plant (though subsequent iterations have much more accurate estimates). 

\begin{figure}
\centering
\begin{overpic}[width=.49\linewidth,trim={2cm 0.5cm 1.4cm 0.3cm},clip]{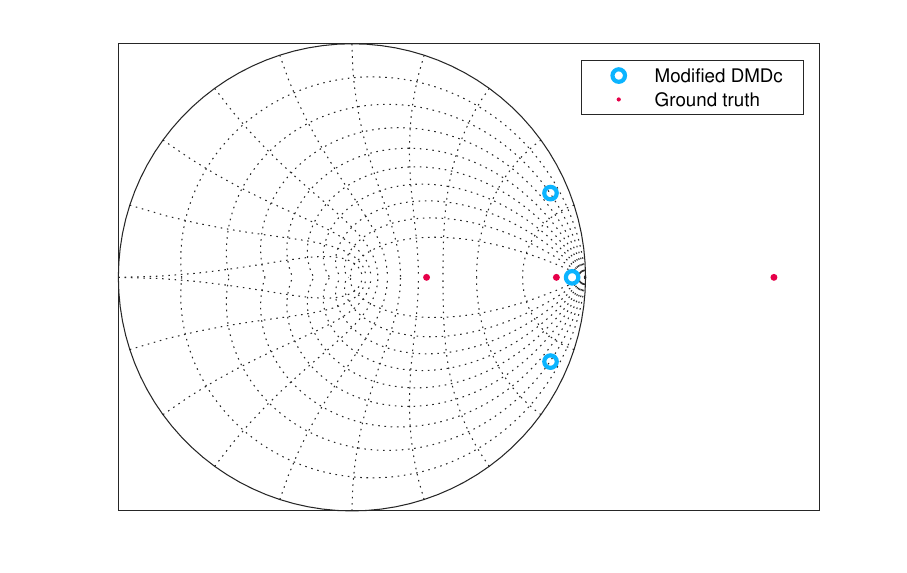}
\put(10, 660){(a)}
\end{overpic}
\begin{overpic}[width=.49\linewidth,trim={2cm 0.5cm 1.4cm 0.3cm},clip]{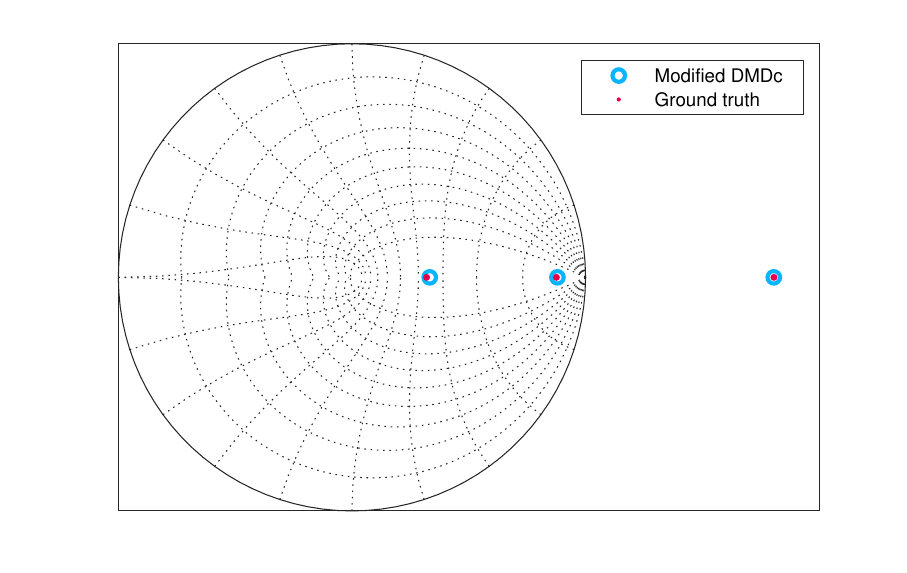}
\put(10, 660){(b)}
\end{overpic}
\caption{Poles of the discrete-time Lorenz system obtained by linearization of the controlled system compared to ground truth positions. (a) At iteration 4, the approximation is still not satisfactory, which is explained by the nonlinear dynamics present in the data employed. (b) At iteration 9, the estimated poles are found considerably closer to the ground truth solution.} 
\label{fig:lorenz_poles}
\end{figure}

Finally, a modal stability analysis is conducted, where poles are computed for the controlled Lorenz system. The modified DMDc approach is employed to data obtained from a single iteration of the \textit{sweep} mode. Similarly to what is observed in figure \ref{fig:lorenz_eq}(b), later iterations tend to provide better approximations of equilibria. Figure \ref{fig:lorenz_poles} presents results for iterations 4 and 9. For comparison, the Lorenz system equations are linearised analytically, and the obtained poles $s_i$ are discretised so that $z_i = e^{s_i \Delta t}$. These poles are presented as ``ground truth'' in figure \ref{fig:lorenz_poles}. As also shown in figure \ref{fig:iterative}, at iteration 4, the perturbed controlled system is driven to nonlinear orbits. At iteration 9, sampled states are concentrated much closer to equilibrium, thus providing better approximations through linear regression.

\subsection{Modified Kuramoto-Sivashinsky equation}

\begin{figure}
\centering
\includegraphics[width=.88\linewidth,trim={0.5cm 0cm 1.0cm 0cm},clip]{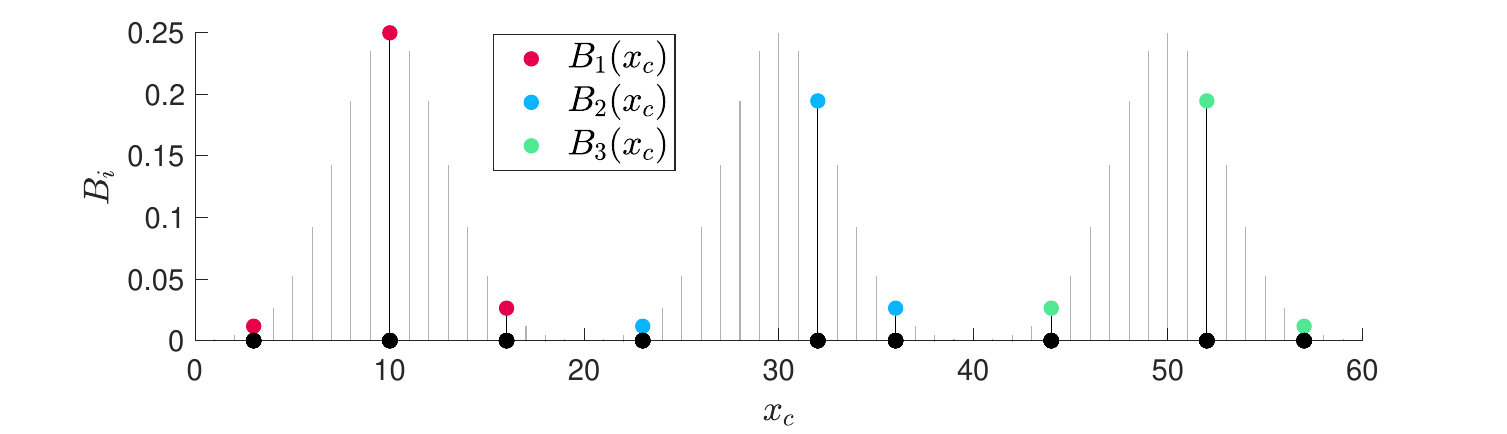}
\caption{Sparse configuration obtained in training for the case with sensor placement, where 9 sensors are kept for the KS equation.}
\label{fig:ks_setup_sparse}
\end{figure}

For the modified KS equation, two different training approaches are presented. They only differ in terms of application of the sparsity layer. One of the cases does not include the sparsity layer on the neural network, (i.e., the full set of sensors is employed), whereas the other case 
includes a sparsity layer to reduce the number of sensors required. In this second case, the total number of sensors used to infer a model for the system dynamics, as well as to conduct closed-loop control, is reduced from 60 to 9, as represented in figure \ref{fig:ks_setup_sparse}. In both cases, the trained process is conducted along 10 iterations of \textit{sweep}, \textit{release} and \textit{control}. We observe in  figure \ref{fig:ks_setup_sparse} that the sparse sensors are selected to be approximately evenly distributed throughout the domain, with three sensors located across each of the three Gaussian functions defining the actuators, and with one sensor at or slightly downstream of each of the actuator peaks.

\begin{figure}
\centering
\begin{overpic}[width=.98\linewidth,trim={1.5cm 0cm 2.7cm 0cm},clip]{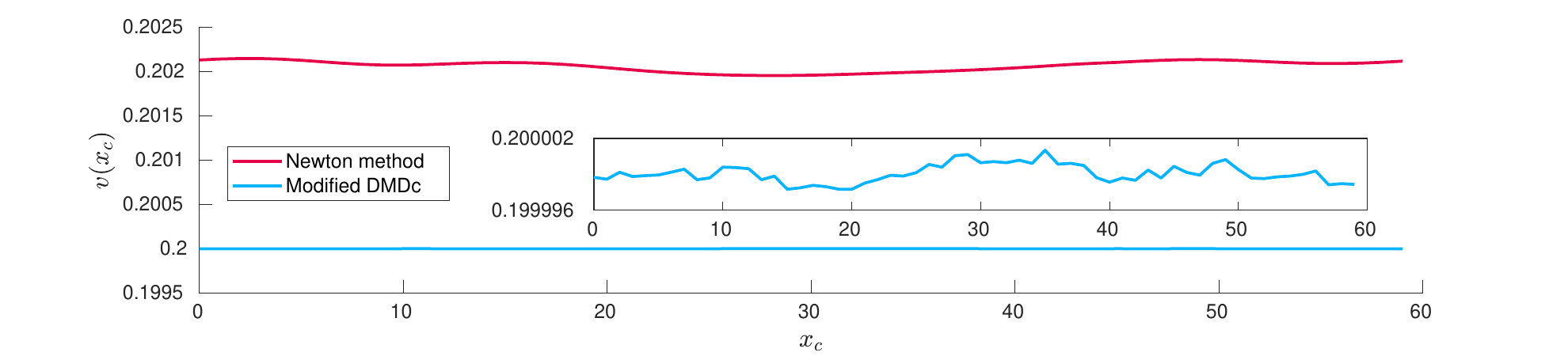}
\put(0, 230){(a)}
\end{overpic}
\begin{overpic}[width=.98\linewidth,trim={1.5cm 0cm 2.7cm 0cm},clip]{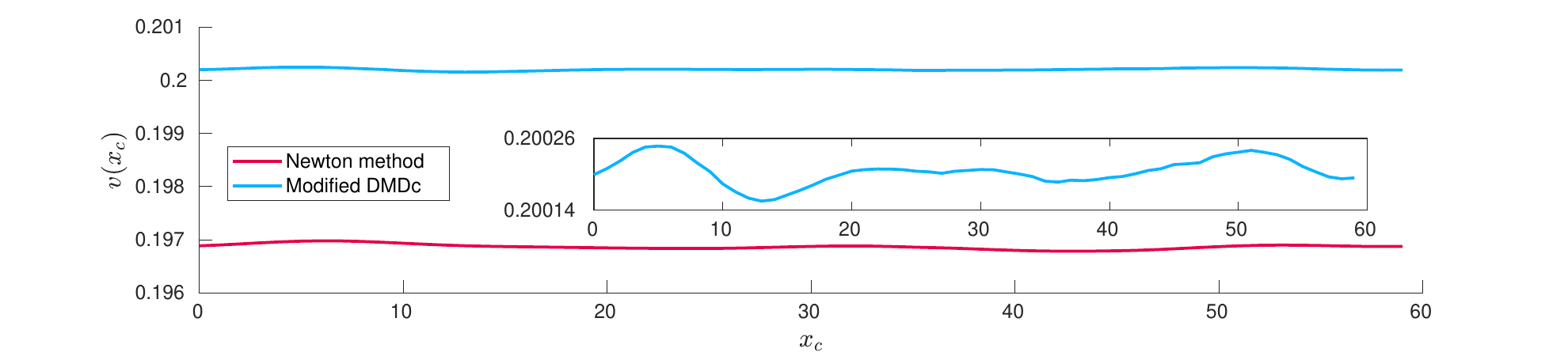}
\put(0, 230){(b)}
\end{overpic}
\caption{Estimated equilibrium points of the KS equation found with (a) full and (b) sparse set of sensors. The true equillibrium is $v(x_c) = 0.2$.}
\label{fig:ks_eq}
\end{figure}

\begin{figure}
\centering
\begin{overpic}[width=.99\linewidth,trim={0.5cm 0cm 1.0cm 0cm},clip]{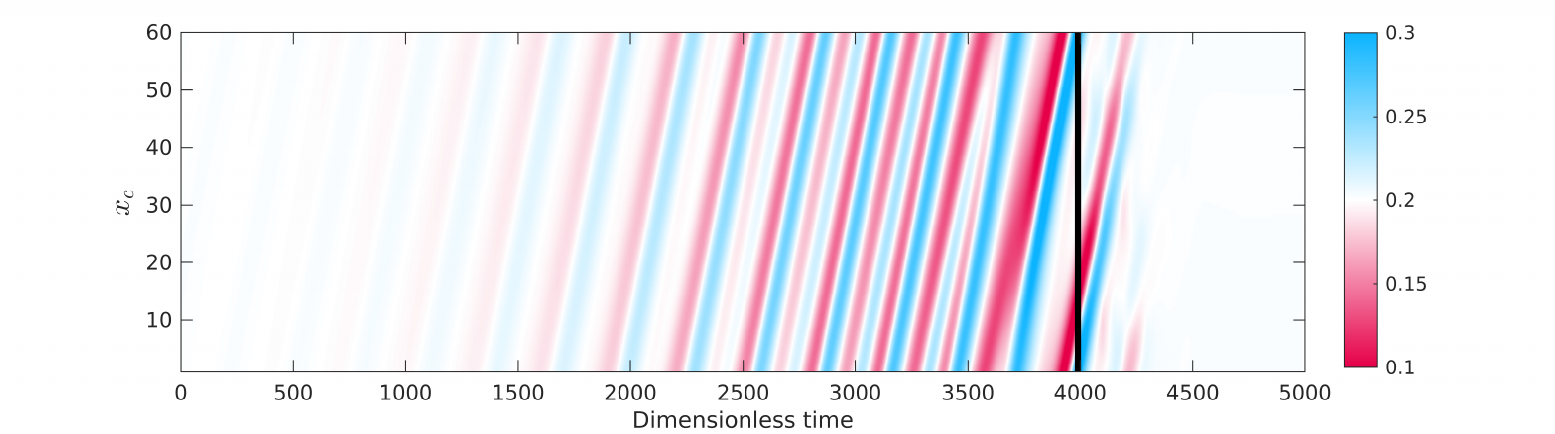}
\put(10, 260){(a)}
\end{overpic}
\begin{overpic}[width=.99\linewidth,trim={0.2cm 0cm 1.0cm 0cm},clip]{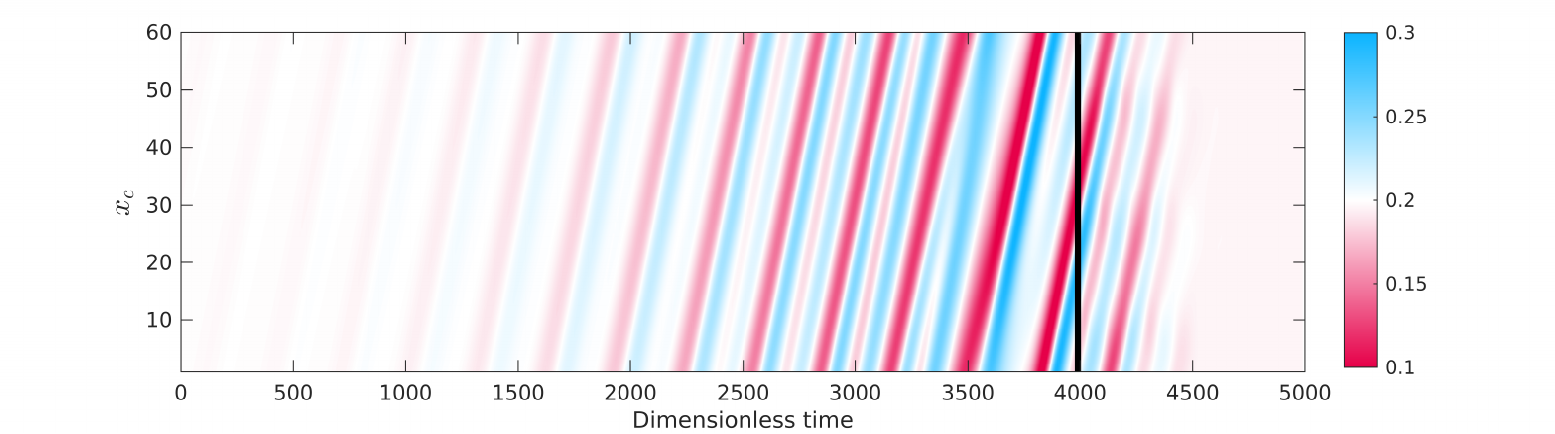}
\put(10, 260){(b)}
\end{overpic}
\caption{Space-time map showing the growth (decay) of disturbances in the KS equation before (after) turning the controller for the (a) full and (b) sparse set of sensors. The black vertical line highlights the instant when the control is turned on.}
\label{fig:ks_spacetime}
\end{figure}

Figure \ref{fig:ks_eq} shows the calculation of the equilibrium points along $x_c$ found in each case using Newton's method and the modified DMDc. It is possible to observe in figure \ref{fig:ks_eq} (a) that the modified DMDc is able to enhance the accuracy of the equilibrium point when the full set of sensors is employed. However, as a trade-off for reducing the number of sensors, figure \ref{fig:ks_eq} (b) shows that the estimate of the equilibrium is worsened with sparse sensors. Despite this, the trained NNC is still able to stabilise the system, as exemplified in figure \ref{fig:ks_spacetime}, where the waves are controlled until convergence.

\begin{figure}
\centering
\begin{overpic}[width=.49\linewidth,trim={2cm 0.5cm 1.4cm 0.3cm},clip]{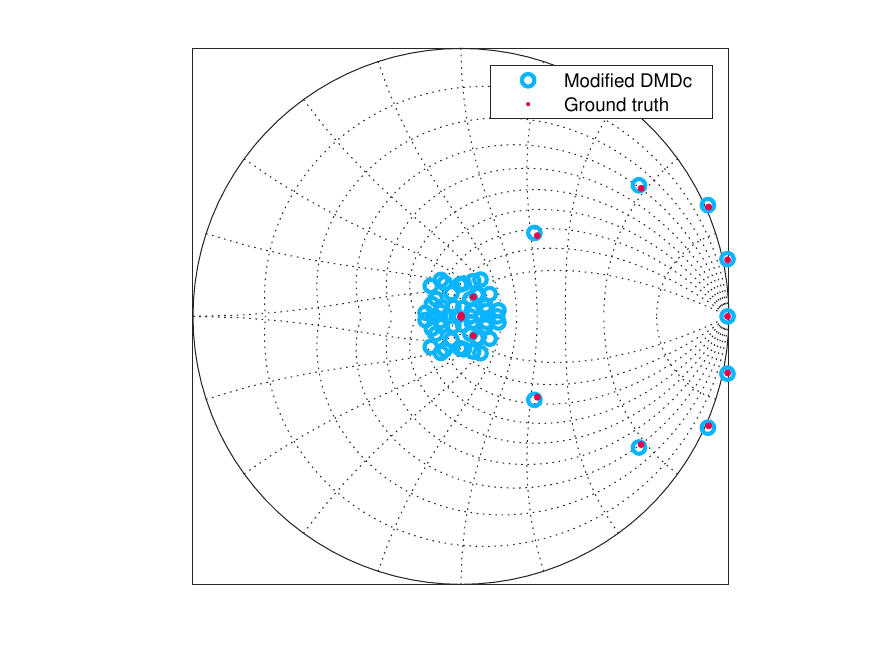}
\put(10, 760){(a)}
\end{overpic}
\begin{overpic}[width=.49\linewidth,trim={2cm 0.5cm 1.4cm 0.3cm},clip]{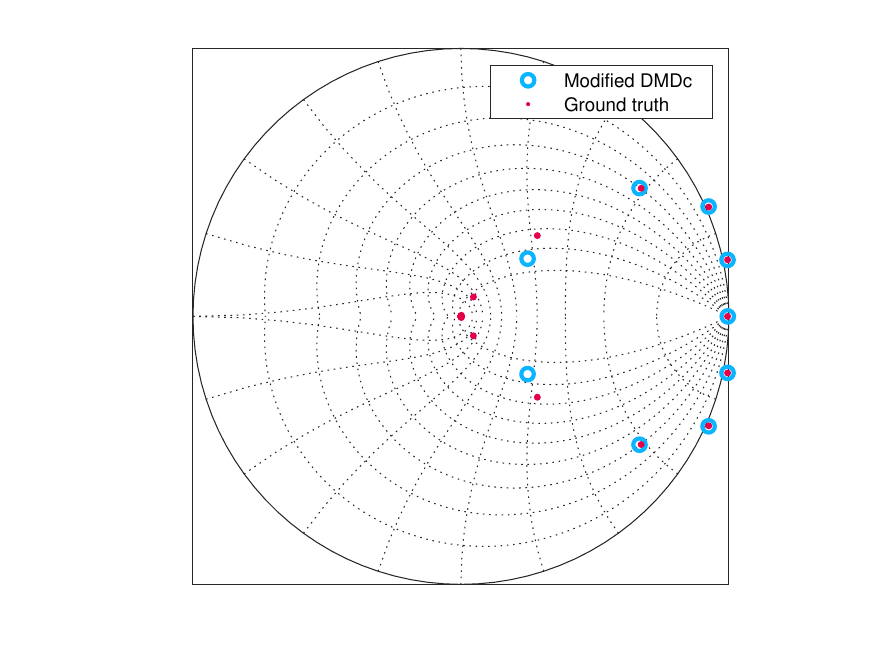}
\put(10, 760){(b)}
\end{overpic}
\caption{Poles of the discrete-time modified KS system obtained by linearization of the controlled system compared to the ground truth solution. The (a) full and (b) sparse sensor approaches are compared. Most of the ground truth poles are concentrated close to the origin, making them harder to visualise.} 
\label{fig:ks_poles}
\end{figure}

For the modal stability analysis of the modified KS equation, the computed poles are shown in figure \ref{fig:ks_poles}. The ground truth poles are obtained by linearizing equation \ref{eq:ks_mod} with a discretization performed using fourth-order centered finite difference schemes for all spatial derivatives. 
The poles are found by computing the eigenvalues $s_i$ of the matrix built from the approximation (with periodic boundary conditions) and transformed to their discrete version $z_i = e^{s_i \Delta t}$. The map obtained with the proposed technique is fairly accurate, with good preservation of frequency and growth rates of the eigenvalues closest to the unit circle. Since the sparse case comprises only 9 sensors, only 9 eigenvalues are found. 

\subsection{Periodic 2D channel flow}
\label{sec:2Dchan}
\begin{figure}
\centering
\begin{overpic}[width=.49\linewidth,trim={1cm 0.5cm 1cm -0.4cm},clip]{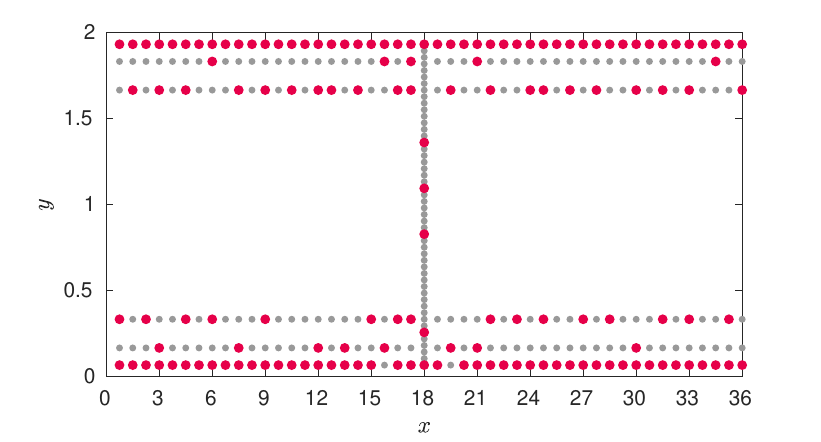}
\put(250, 560){Horizontal velocity probes}
\end{overpic}
\begin{overpic}[width=.49\linewidth,trim={1cm 0.5cm 1cm -0.4cm},clip]{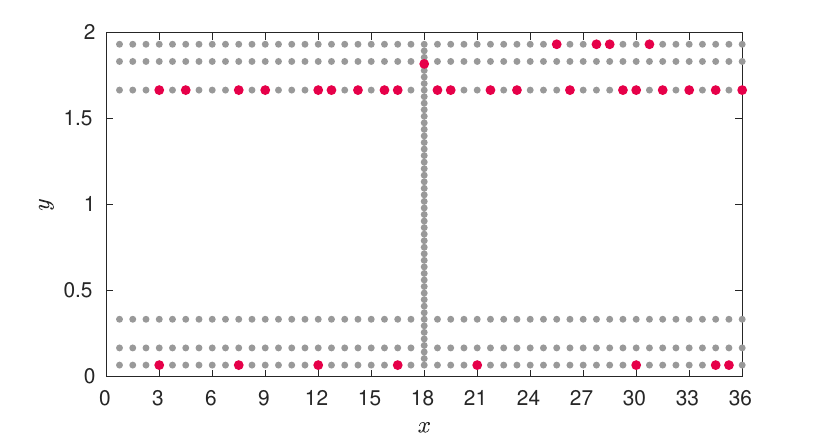}
\put(280, 560){Vertical velocity probes}
\end{overpic}
\caption{Probe locations employed with sparse sensing, where the number of probes is reduced from a total of 664 to 181, from which 148 are used for the horizontal velocity component, while 33 are used for the vertical one. The grey dots are deactivated by the sparsity layer.}
\label{fig:channel_probes}
\end{figure}

The 2D channel flow case is studied with dynamic sensor placement. The final distribution of probes after 19 iterations of training (with \textit{sweep}, \textit{release}, and \textit{control} modes) is presented in figure \ref{fig:channel_probes}. It is possible to notice the preference for horizontal velocity probes ($u$) rather than vertical ($v$) ones. The sensor placement is able to achieve a reduction from 664 to 181 signals, 148 and 33 correspond to $u$ and $v$ velocity probes, respectively. The spatial distribution of these chosen sensors shows that almost all of the near-wall $u$ sensors are included, possibly highlighting the importance of near-wall information for estimating and controlling (with wall-based actuation) the system dynamics.

\begin{figure}
\begin{overpic}[width=.96\linewidth,trim={0cm 0cm 0cm -4.5cm},clip]{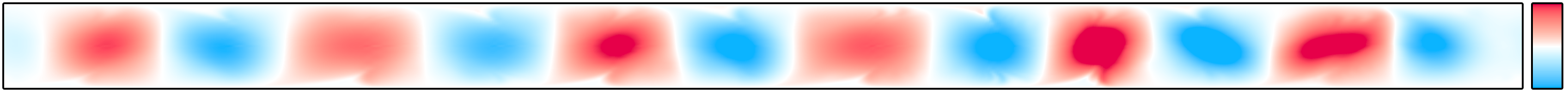}
\put(448, 64){\fontsize{8}{8}\selectfont Uncontrolled}
\put(1000,3){\fontsize{6}{8}\selectfont -2e-1}
\put(1000,44){\fontsize{6}{8}\selectfont +2e-1}
\end{overpic}
\begin{overpic}[width=.96\linewidth,trim={0cm 0cm 0cm -4.5cm},clip]{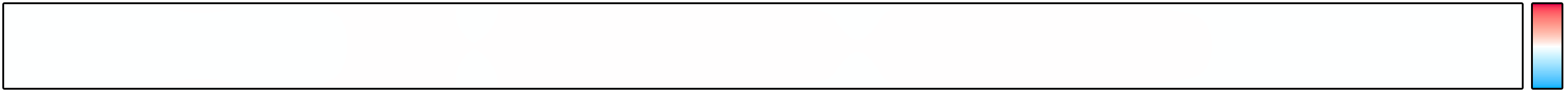}
\put(461, 64){\fontsize{8}{8}\selectfont Controlled}
\put(1000,3){\fontsize{6}{8}\selectfont -2e-1}
\put(1000,44){\fontsize{6}{8}\selectfont +2e-1}
\end{overpic}
\begin{overpic}[width=.96\linewidth,trim={0cm 0cm 0cm 0cm},clip]{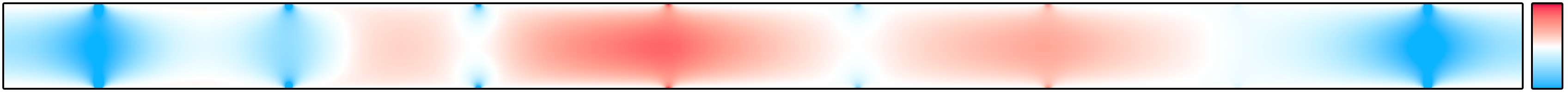}
\put(1000,3){\fontsize{6}{8}\selectfont -5e-4}
\put(1000,44){\fontsize{6}{8}\selectfont +5e-4}
\end{overpic}
\caption{Comparison of uncontrolled and controlled flows through visualization of $v$-velocity contours (see Movie 1). For the controlled case, a more saturated contour range is also shown in the bottom plot, and one can see that the actuation does not converge to zero due to an imperfect estimation of the equilibrium.}
\label{fig:channel_control}
\end{figure}
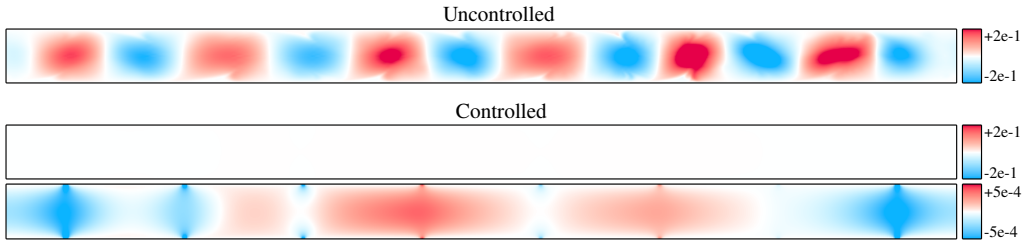

With feedback of the signals probed at the locations obtained through training, the NNC is able to completely stabilise the flow. A comparison of the uncontrolled and controlled flows is presented in figure \ref{fig:channel_control}, where $v$-velocity contours are shown. As expected, the actuation and the flow field do not converge to the exact equilibrium, since its estimate is not perfect, especially considering that, for control, we employ the approximation obtained through Newton's method (see section \ref{sec:stability}). The controlled flow is presented in two ways: 1) using the same contour range of the uncontrolled flow, and 2) with a more saturated contour range that allows one to notice some small residual velocity fluctuations. When employing the same scale as that of the uncontrolled case, the bias becomes nearly imperceptible. For the uncontrolled case, the snapshot represents a frame captured from an unsteady flow, where the flow structures are transported from left to right, while the controlled case is simply a steady state. 

The present flow includes two unstable modes. These are computed through the eigendecomposition of the associated Orr-Sommerfeld operator, which is discretised using a pseudo-spectral method with Chebyshev polynomials utilizing the toolbox by \citet{weideman2000matlab}. The wavelengths of the unstable modes obtained by the Orr-Sommerfeld equation correspond to 1/6 (mode 1) and 1/5 (mode 2) of the channel length. This is in agreement with the wavelengths of the dominant structures observed in simulations of the uncontrolled nonlinear system, as shown in figure \ref{fig:channel_control}(a).

\begin{figure}
\centering
\begin{overpic}[width=.49\linewidth,trim={1cm 0.5cm 1cm -0.4cm},clip]{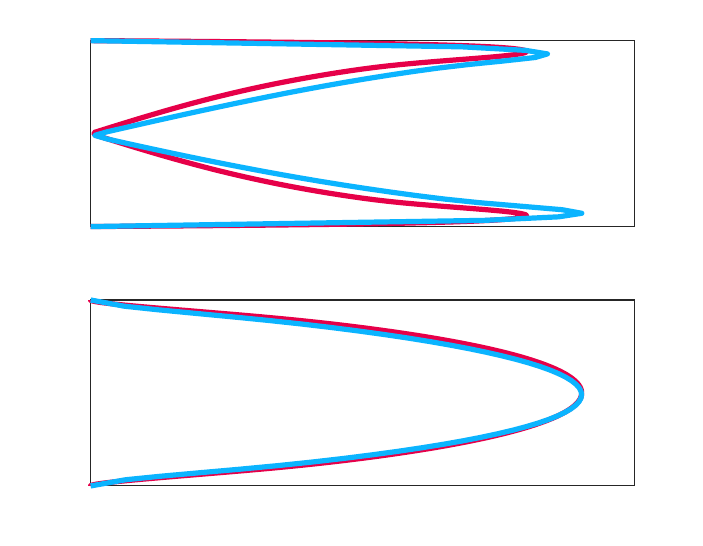}
\put(0, 493){\rotatebox{90}{\fontsize{7}{8}\selectfont Horizontal velocity}}
\put(0, 078){\rotatebox{90}{\fontsize{7}{8}\selectfont Vertical velocity}}
\put(450, 828){\fontsize{7}{8}\selectfont Mode 1}
\end{overpic}
\begin{overpic}[width=.49\linewidth,trim={1cm 0.5cm 1cm -0.4cm},clip]{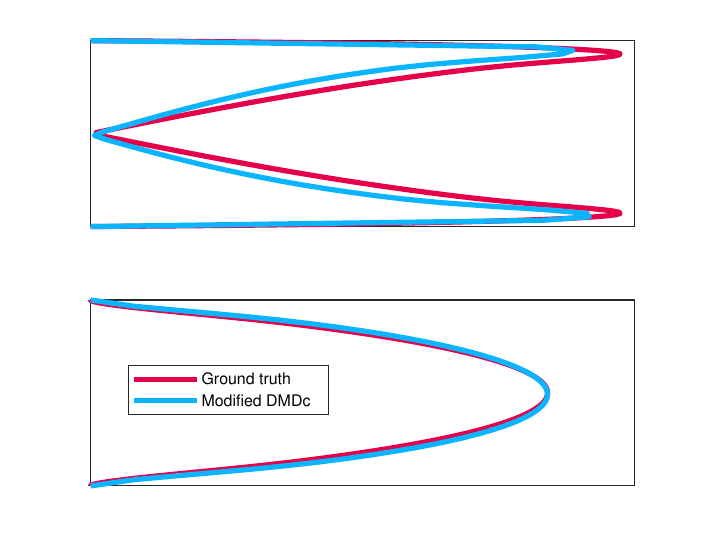}
\put(450, 828){\fontsize{7}{8}\selectfont Mode 2}
\end{overpic}
\caption{Unstable modes for the 2D channel flow. Ground truth and estimated results are compared in terms of absolute values for the eigenvectors mapped to the vertical line of sensors.} 
\label{fig:channel_modes}
\end{figure}

By applying the modified DMDc approach to the data obtained near equilibrium,  two unstable pairs of modes are found, corresponding to the two unstable modes of the true linearised system. For this case, the discrete poles $z_i$ found with the modified DMDc are converted to a continuous version $s_i=\ln{(z_i)}/\Delta t$, so that one can directly infer the corresponding growth rates and frequencies. A comparison between the true and estimated unstable eigenvalues is shown in Table \ref{tab:mode_poles}, where close agreement is observed.
In these results, the imaginary part corresponds to the oscillation frequency while the real component provides the growth rate. The higher relative error found for the growth rate when compared to the frequency can be explained by the transient time scales being too slow compared to the discretization time $\Delta t$. Since $\Delta t$ needs to be small enough to capture the signals without significant aliasing, and the timescale of the oscillations are considerably faster than the growth rate, the discrete time horizon to capture the long-term growth becomes too large. The present approach is also able to provide the associated eigenvectors. In figure \ref{fig:channel_modes}, the absolute values of the $u$ and $v$-velocity eigenvectors related to each pair of eigenvalues are plotted. Here, these eigenvectors are mapped to the vertical line of sensors for comparing with the ground truth values. Despite displaying a slight asymmetry for the proposed method, which could likely be improved by adding more sensor points, the comparisons are in good agreement. 

\begin{table}
  \begin{center}
\def~{\hphantom{0}}
  \begin{tabular}{lccc}
         & Mode 1 & Mode 2\\[3pt]
        Ground truth & $1.8601$e-3 $ \pm j2.6403${e-1 } & $6.8185${e-4} $ \pm j2.0206${e-1} \\
        Modified DMDc & $2.0545${e-3} $\pm j2.6396${e-1} & $7.4956${e-4} $\pm j2.0214${e-1} \\
  \end{tabular}
  \caption{Eigenvalues for unstable modes 1 and 2 for 2D channel flow. Colours correspond to modes plotted in figure \ref{fig:channel_modes}.}
  \label{tab:mode_poles}
  \end{center}
\end{table}

\subsection{Confined cylinder flow}
\begin{figure}
\centering
\begin{overpic}[width=.49\linewidth,trim={1.3cm 1.5cm 1cm 0.0cm},clip]{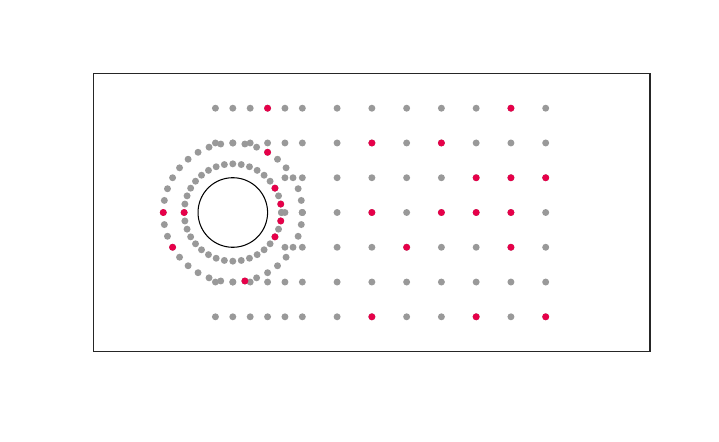}
\put(250, 515){Horizontal velocity probes}
\end{overpic}
\begin{overpic}[width=.49\linewidth,trim={1.3cm 1.5cm 1cm 0.0cm},clip]{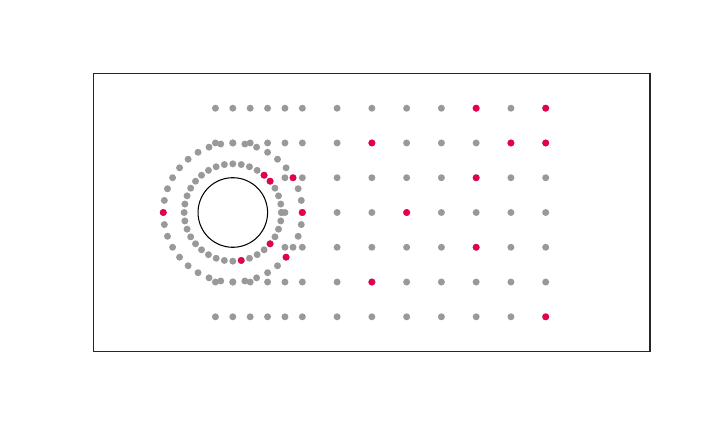}
\put(280, 515){Vertical velocity probes}
\end{overpic}
\caption{Demonstration of the application of the sparsity layer for the confined cylinder flow. The total number of measurements is reduced from 306 to 47 (25 and 18 probes for the horizontal and vertical velocity components, respectively). The grey dots are deactivated by the present L1 regularization.} 
\label{fig:cylinder_probes}
\end{figure}

\begin{figure}
\centering
\includegraphics[width=.99\linewidth,trim={1.9cm 0cm 2.1cm 0cm},clip]{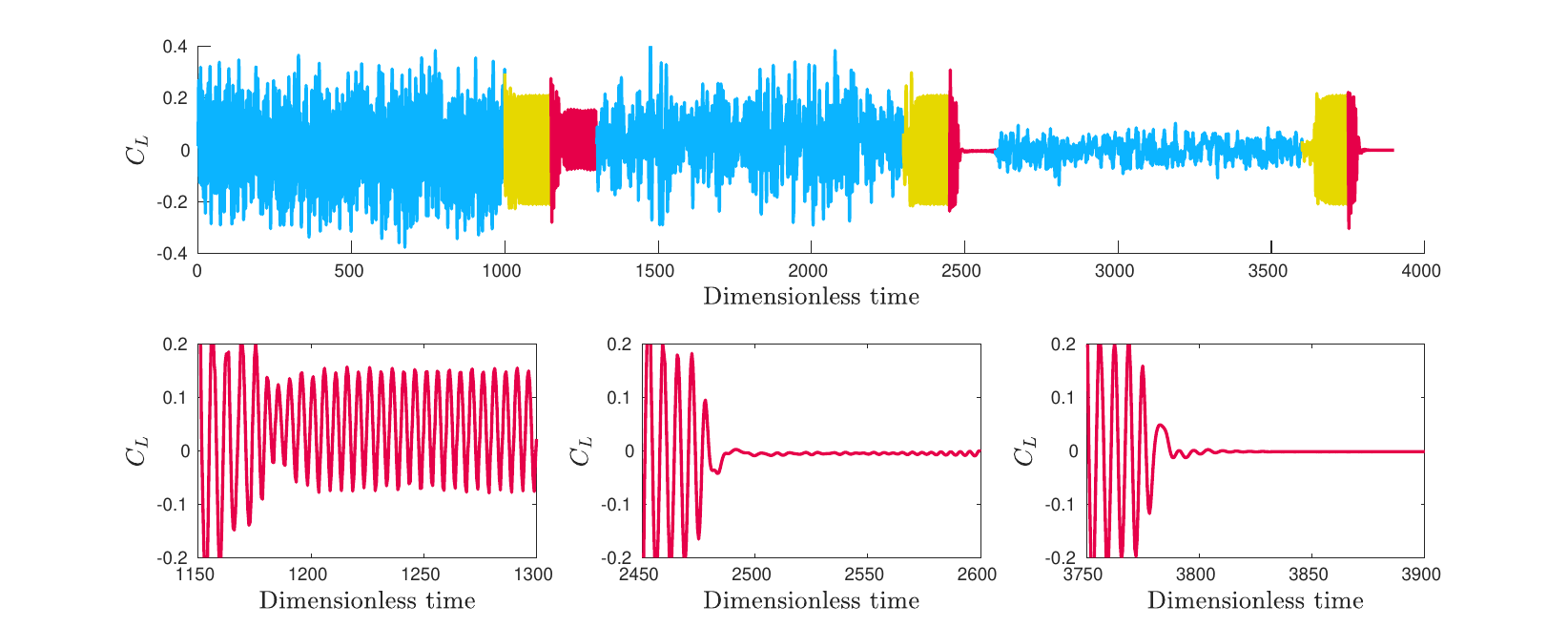}
\caption{Evolution of lift coefficient for 3 iterations of the \textit{sweep} (blue), \textit{release} (yellow), and \textit{control} (red) stages for the confined cylinder flow.}
\label{fig:cylinder_signals}
\end{figure}

As was the case in section \ref{sec:2Dchan}, this system is modelled and controlled while also incorporating a sparsity layer in the NNSM to select sensors. In doing so, the NNSM training process for the confined cylinder flow selects a total of 47 sensors from the candidate set of 306 sensors: 25 and 18 points for $u$ and $v$ velocity components, respectively. While the number of sensors is significantly reduced, as presented in figure \ref{fig:cylinder_probes}, the NNC approach with iterative training is able to successfully stabilise the this flow. To provide an overview of the training process, the lift coefficient over the cylinder is presented in figure \ref{fig:cylinder_signals} for the first 3 iterations. 
In the first iteration, the NNSM and the NNC are trained using only open-loop data. For this iteration, the lift fluctuations are only slightly reduced when flow control is turned on. For the following iterations, the models are trained using both open and closed-loop data, which improves the results, as more data is available near the equilibrium point, where the system is approximately linear. 

While the stabilization is not completely achieved by the end of the second iteration, 
a reduction in the main oscillations can be observed. By the third iteration, complete stabilization is achieved through control. 

\begin{figure}
\centering
\includegraphics[width=.99\linewidth,trim={1.3cm 0cm 1.5cm 0cm},clip]{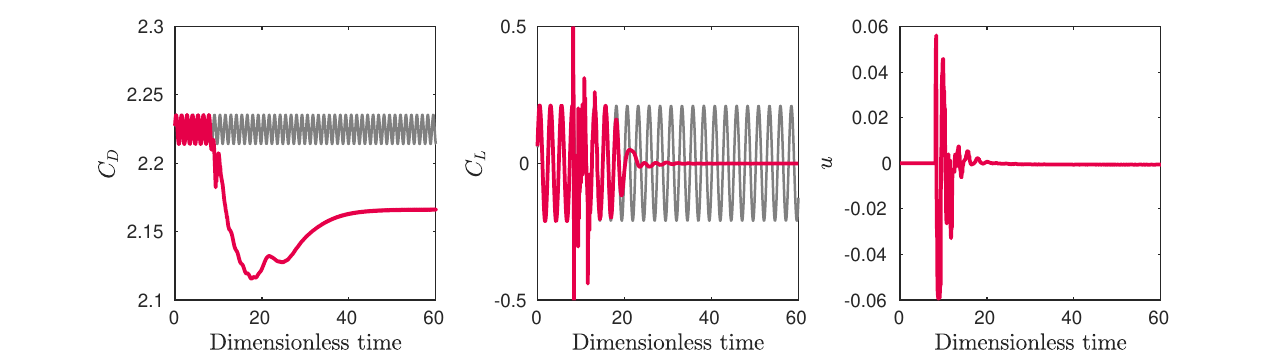}
\caption{History of drag and lift coefficients, and control effort for the third iteration in the control mode. Gray lines show uncontrolled solution.}
\label{fig:cylinder_cx}
\end{figure}

\begin{figure}
\begin{overpic}[width=.999\linewidth,trim={0cm 0cm 0cm -3cm},clip]{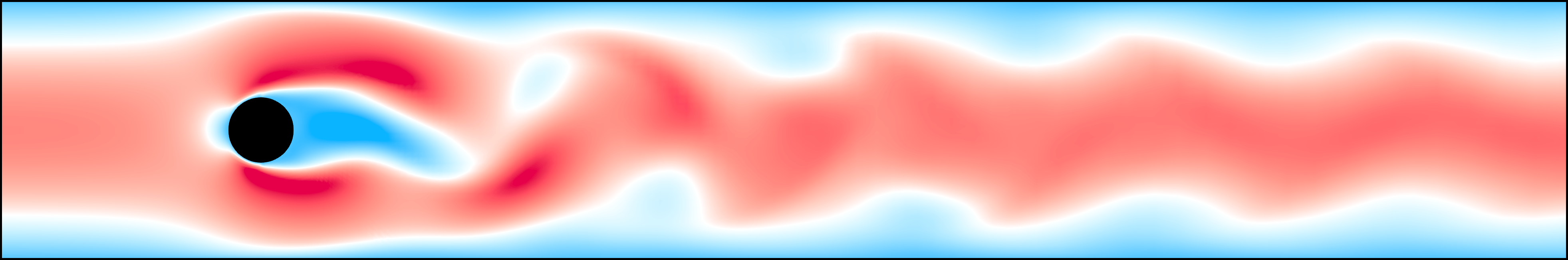}
\put(448, 169){\fontsize{8}{8}\selectfont Uncontrolled}
\end{overpic}
\begin{overpic}[width=.999\linewidth,trim={0cm 0cm 0cm -3cm},clip]{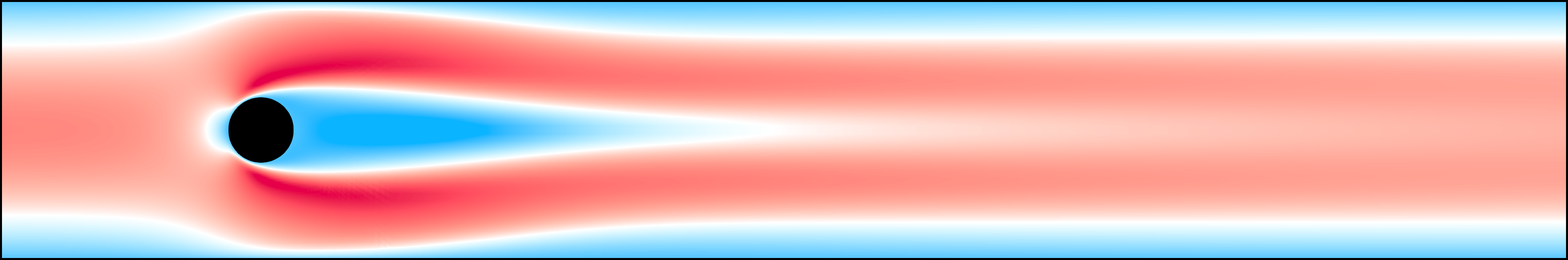}
\put(461, 169){\fontsize{8}{8}\selectfont Controlled}
\end{overpic}
\caption{Comparison of uncontrolled and controlled $u$-velocity fields (see Movie 2). The former is represented by a snapshot taken at the flow natural limit cycle. For the controlled case, the snapshot is captured after stabilization is achieved.}
\label{fig:cylinder_control}
\end{figure}

Important features observed in the controlled flow are shown in figure \ref{fig:cylinder_cx}, which shows the convergence history of the drag and lift coefficients, as well as of the control input, during the control stage of the third iteration. The present flow control approach is able to suppress both the drag and lift coefficient fluctuations. At the same time, the mean drag is also reduced. The complete stabilization also ensures that only a minimum effort is required to keep the system operating with small oscillations and drag losses. Residual efforts are due to imperfect estimation of equilibrium, and may also be required for compensating eventual perturbations. Figure \ref{fig:cylinder_control} shows $u-$velocity contours for the uncontrolled and controlled cases. In the former, vortex shedding develops along the wake, inside the plane channel. On the other hand, a steady solution is obtained for the controlled case.

To contextualize these observations, we now discuss prior control results for this confined cylinder configuration. The most direct comparison can be made with \citet{li2022reinforcement}, who also attempted to stabilize the flow.
They proposed an approach for the stabilization problem using reinforcement learning to train neural networks capable of minimizing the vortex shedding energy, which is translated to a reward function that takes the velocity fluctuations at selected probe locations.
Such calculations could be done by either subtracting the mean flow or the equilibrium flow found through selective frequency damping (SFD) from the measured velocity probes. By using the former approach, convergence was not observed, whereas using the equilibrium states from SFD enabled stabilization. The results, however, required manual positioning of probes through a heuristic approach informed by considering the wavemaker region obtained from considering the region of overlap between the leading direct and adjoint eigenmodes. Furthermore, the prior computation of the equilibrium through SFD is a dedicated step that requires additional simulation-intrusive computations. 
By contrast, our approach enables automatic equilibrium (and eigenmode) computations, which do not require the use of modified numerical solvers.
\citet{li2022reinforcement} found that successful control of the confined cylinder flow resulted in a small, nonzero mean control signal (i.e., control jet flowrate), which we also observe here. In our case, this is on the order of $\bar{u} \approx $ -6e-4, which is too small to observe on the right subplot of figure \ref{fig:cylinder_cx}.

It is more difficult to make a direct comparison with other prior control
studies of this system, primarily due to the different control objectives considered. The current work aims for stabilization, whereas other prior investigations considering this confined cylinder flow \citep{rabault2019artificial, varela2022deep} focused on drag minimization. As we holistically approach the control problem proposing methods that also provide estimates of the equilibrium, the stabilization task is directly enabled through NNC training. For this confined cylinder flow, a drag reduction is seen through stabilization,  the lift and drag oscillations are suppressed, and the control input required to keep the controlled system at the unstable equilibrium approaches zero. 
While we find a drag reduction of 2.6\% for the fully-stabilized flow, larger drag reduction---5.7\% at $\mathrm{Re}=100$ and 21.6\% at $\mathrm{Re}=200$---are reported in \citet{varela2022deep}. This suggests that the optimal solution for drag reduction is not the flow at its equilibrium. This can be verified, for example, in the strong overshoot seen in $C_D$ (Fig. \ref{fig:cylinder_cx}), where the maximum drag reduction is reached before the flow is stabilized. In fact, in iteration 1, the control system produces an average drag reduction of 3.4\%, lower than the stabilized flow at iteration 3 of 2.6\%.

\section{Conclusions} \label{sec:conclusions}

In the present work, neural networks are applied to perform as surrogate models of nonlinear dynamical systems, and also as controllers for stabilizing such systems. The proposed approach is first tested with the Lorenz set of ordinary differential equations, as well as with a modified version of the Kuramoto-Sivashinsky partial differential equation. Then, flow control is demonstrated for a periodic 2D channel flow, for which a data-driven linear stability analysis is also conducted and, finally, for a more complex case of a confined cylinder flow.  
The trained neural networks provide adequate surrogate models that achieve all the present results with a single hidden layer, except for the confined cylinder case, which employs two layers.
In the same fashion, the neural networks obtained to control the investigated systems are also single-layer models with only a few neurons. Despite the simple individual neural network architectures, through a recurrent training strategy, the present framework is able to obtain effective controllers.  
Results also support the efficacy of neural networks for learning complex system dynamics, as well as to obtain important flow features, such as equilibrium points, and leading eigenmodes of the linearised system about these fixed points. Although these data-driven models are trained as black boxes, 
relevant information can be extracted by using backpropagation.

In this work, several different nonlinear unstable systems are investigated, and their dynamics do not naturally converge to equilibrium points.
This behaviour hinders data sampling that represents the plant dynamics close to such points, where linear approximations should be able to inform the main eigenvalues and linear modes. As a solution for this problem, we propose the application of NNC with an iterative training of the models. {As well as achieving stablization,} this new method proves to be an efficient approach for equilibrium computation, sensor placement, and modal stability analysis. As demonstrated by results, the iterative approach is key for bringing the system closer to a progressively better estimation of the equilibrium point. In all the cases presented, the first iteration is unable to provide accurate results, specially for equilibrium computation and modal analysis. Even for the control task, stabilization is not achieved, for example, in the cylinder flow case. The improvement achieved with the iterative process puts the presented NN approach as an efficient and accurate candidate for flow stabilization and data-driven flow analysis, with the possibility of enabling limited sensor allocation.

The first plant studied with the proposed methodology is the Lorenz system. With the chosen numerical parameters, its dynamics results in a chaotic attractor featuring 3 different equilibrium points. 
While not explicitly shown, it was found that the initial guess for the equilibrium point  
has an important role in the iterative process, which means in general that if initial guesses are too far from an equilibrium of interest, the algorithm may converge to another point, which might not be desirable.
In the case presented in the results section, the initial choice at $x=y=z=0.1$ converges near the origin after a few iterations. Results from this first test indicate that the iterative training of the NNSM and NNC successfully overcomes the lack of data sampled near the equilibrium. As the control strategy improves and the perturbations are reduced in amplitude, datasets containing more measurements near equilibrium are built, improving the quality of linear approximations and the control design aiming stabilization. Linear regression through DMD-based algorithms is also enabled for a considerably broader 
range of plants, taking advantage of data produced by controlled nonlinear systems that are brought closer to linear operation through feedback control. The modification to the DMDc method used in this work also helps extend the analyses to systems whose equilibrium points are unknown or poorly estimated, allowing for a model fitting that also enhances estimates of equilibrium bias.

The present methodology is next applied to control a more complex problem,  
a modified version of the Kuramoto-Sivashinsky equation. 
This modified equation is proposed in order to force a single equilibrium point instead of a continuous space of possibilities. This test case allows the study of the effectiveness of sensor placement through L1 regularization of the inputs. Again, equilibrium estimation and feedback control 
obtain satisfactory results, even with a reduction in the number of sensors from 60 to 9. Although this reduction slightly worsens the estimation of equilibrium and, therefore, adds further errors to control convergence, tuning the L1 regularization through the choice of a single hyperparameter leads to a configuration with an intermediate number of sensors at reduced accuracy cost. Even so, the application of Newton's method with linearization of the NNSM through backpropagation is able to provide an equilibrium point estimation that allows for stabilization and, therefore, to obtain the approximately linear dataset subsequently used to identify the dominant eigenmodes of the linearized system.

Application of the NNC is then applied to the Navier-Stokes equations. As a first case, stabilization of a 2D streamwise-periodic channel flow is sought, where the capability of conducting a linear stability analysis is also explored for a flow configuration with two unstable modes. The L1 regularization is applied in this case and, from a total of 664 measured signals, training is able to reduce sensing to 181 probes. Stabilization is achieved with 8 pairs of actuators in opposition through NNC, which enables sampling of data close to equilibrium, i.e., of signals described by approximately linear dynamics. The data is used for performing a linear regression through a modified DMDc. The unstable eigenvalues and eigenmodes computed by this technique show good agreement with those obtained directly from the solution of the linear Orr-Sommerfeld operator.
The different magnitudes of the growth rate and frequency of the unstable modes result in different levels of accuracy for their estimation, with the growth rates being close to 0. In this sense, the frequencies present lower relative errors than the growth rates due to the large discrete time intervals associated with the latter for the chosen time step. The natural growth rate corresponds to a very small component of the measured signal compared to the effect of the uncontrolled frequency.

Finally, control of a confined cylinder flow is also presented. The setup chosen with actuators in opposition is the same proposed in different studies found in literature \citep{rabault2019artificial,li2022reinforcement}. Here, we assess the ability of the neural network model to stabilise a complex system through the iterative training framework. Even for this more complex case, the NNC is built from a single layer containing only 8 nodes. Results show that flow control suppresses the lift and drag oscillations from vortex shedding and also reduces the mean drag by bringing the flow to its equilibrium point. Not only is the system  successfully controlled, but additionally the low complexity of the NNC suggests that there is considerable room for applying the technique to considerably more complex fluid systems without prohibitive costs. For this confined cylinder configuration, prior work \citep{li2022reinforcement} found that RL control performance could be improved through equilibrium computation and linear stability analysis in the vicinity of this equilibrium. Here, we achieve similarly successful control results without needing to perform these auxiliary computations explicitly, but where the results of such computations are available as a byproduct of our modeling and control methodology.

In each of the examples featuring partial differential equations across a spatial domain, we have utilized a sparsity layer in the NNSM to reduce the number of sensor measurements (and thus model inputs). While the resulting sensor locations are chosen to optimise a given cost function, they are not necessarily unique solutions, in the sense that alternative sensor locations may be obtained for different realizations (with different random input signals and randomly initialized NN weights) of the training procedures. This is clear from observing that the identified sensors do not follow the same symmetry properties as the geometries of the problem. That being said, the chosen sensor locations do reveal some aspects of the underlying physics. For the modified KS equation, the sensors are distributed throughout the domain, in accordance to the translation invariance of the problem (aside from the actuator locations). For the 2D channel flow, the sensor locations indicate the importance of near-wall streamwise velocity measurements, with a sparser set of measurements away from the wall. The asymmetry in the chosen locations of the wall-normal velocity probe locations is perhaps due to the fact that the instabilities in this system consist of mode shapes where the vertical velocity component extends across the whole domain, so taking measurements at both the upper and lower walls may be unnecessary. In the confined cylinder flow, we find that sensors both close to the cylinder, and downstream in the wake are chosen, indicating the importance of measurements in both regions for suppressing the natural vortex shedding behavior. While beyond the scope of the present work, it could be interesting to compare the identified sensor locations with those obtained from alternative methods \citep{manohar2018data,sashittal2021data,williams2022data,graff2023information}.

The modeling and control framework utilised here is similar in principle to that used in classical linear control theory, with separate models for the plant and controller connected in feedback. While nonlinear control is substantially more complex, here we show that relatively simple neural networks can be used to develop effective nonlinear models for both the plant and controller, when trained with a method that promotes the generation of large quantities of near-equilibrium data. Future work could also utilise linearization of the plant and/or controller models for linear control design, with better authority over the plant behavior through well studied control design techniques that can ensure robustness, optimality or specific pole placement.

While it was demonstrated that the proposed methodology could be used to identify modal linear amplification mechanisms, future work could extend these methods to study nonmodal amplification, such as transient growth and resolvent analysis. Data-driven implementations of such analyses have been implemented previously \citep{herrmann2021data}, though for non-normal systems it may be more difficult to obtain accurate results without adjoint operators/data, which may require further modifications to the methodology presented here.

We lastly emphasise  that the methods and results are obtained from non-intrusive methods in the sense that only the nonlinear system (with appropriate control inputs) has been used. This means that the methodology could in principle be applied (for both control and flow physics analysis) in an experimental setting. This, as well as the application to systems presenting higher degree of complexity, remains the subject of future work.



\section*{Acknowledgments}
The authors acknowledge the financial support received from Fundação de Amparo à Pesquisa do Estado de São Paulo, FAPESP, under Grants No. 2013/08293-7 and 2021/06448-0. The first author is supported by  FAPESP PhD scholarships No. 2019/19179-7 and 2022/00469-8, which are also acknowledged. STMD acknowledges support from NSF Award  2238770. The computational resources used in this work were provided by CENAPAD-SP (Project 551), and by LNCC via the SDumont cluster (Project SimTurb). 

\section*{Declaration of Interests}
The authors report no conflict of interest.

\appendix
\section{Hyperparameters}\label{appA}

In this appendix, the chosen values for the hyperparameters in each case are presented. They are organised in table \ref{tab:hp}. Regarding the iterative training, the ``number of iterations'' correspond to how many times the proposed training steps are conducted. It is expected that at each iteration the perturbed plant controlled by NNC tend to produce more data near equilibrium. The ``number of steps in \textit{sweep}", ``number of steps in \textit{release}" and ``number of steps in \textit{control}" correspond to the number of measurements sampled in each mode at each iteration. 

 The open loop control signal is parameterised by a ``control input saturation value", which defines the possible uniform interval of random values it can assume. The staircase signal has a step width given by the ``number of time steps for each stair step (open loop signal)". Also, the ``maximum amplitude decay at each iteration" parameter makes sure the open-loop signal is reduced in amplitude at each training iteration.

 From each mode (\textit{sweep}, \textit{release} and \textit{control}), data is sampled for training, but the dataset size is limited to a ``maximum number of points sampled for training NNSM". This ensures that the training cost does not increase beyond a certain point. A ``NNSM learning rate" is defined for the ADAM 
 optimization algorithm, and the model is updated through a ``number of epochs at each training iteration for the NNSM". The ``number of neurons at NNSM hidden layers" hyperparameter defines the number of layers, which is given by the number of elements in the array of numbers, and the number of neurons in each layer, which is the value at each position. Note that, except for the cylinder case in table \ref{tab:hp}, all NNSM models are trained with a single layer. For assessment of possible overfitting, an ``NNSM training set ratio" as a fraction of sampled data is also chosen, which allows for the comparison of losses between a training set and a testing set of samples. The ``hidden layers L2 regularization" parameter is the $r_2$ variable defined in this work. ``Use sparsity layer" is a Boolean that determines if sparse sensor placement is used. If so, we can set the ``sparsity layer L1 regularization" parameter ($r_1$) and the ``sparsity layer truncation tolerance" that determines the absolute value below which the input layer weights are truncated to zero. The input layer is organised according to the ``number of control inputs" and the the ``number of states" measured.

 Similarly a ``maximum number of points sampled for training NNC" is also defined to avoid excessive complexity, as well as the ``NNC learning rate" for ADAM and the ``number of epochs at each training step for NNC". The ``training finite horizon length" in number of steps ($n_h$) is also presented. For the NNC, an ``NNC training set ratio" as a fraction of sampled data is also chosen. The ``number of neurons at NNC hidden layers" is always the same in all cases, which consists of a single layer containing 8 neurons. ``NNC control saturation" stands for the limits of control effort that NNC can provide. It is implemented as an output layer consisting of weighted sigmoid functions. The ``control input weight in loss function" corresponds to $\mathsfbi{w_u}$.

 Finally, regarding equilibrium computation, the ``initial guess for equilibrium point" is presented for each case, corresponding to the first guess used in the first training iteration. At each of these iterations, the estimation of equilibrium is updated along a ``Number of steps for the Newton method".

\begin{table}
  \begin{center}
\def~{\hphantom{0}}
\resizebox{0.8\textwidth}{!}{
\rotatebox{90}{
  \begin{tabular}{rlllll}
        Description & Lorenz & KS - full sensing & KS - sparse sensing & Channel flow & Cylinder flow \\[10pt]
        Number of training iterations & 25 & 10 & 10 & 19 & 3 \\[5pt]
        Number of steps in \textit{sweep} & 1000 & 2000 & 2000 & 1800 & 2000 \\[5pt]
        Number of steps in \textit{release} & 80 & 400 & 400 & 300 & 300 \\[5pt]
        Number of steps in \textit{closed} & 80 & 100 & 100 & 900 & 300 \\[5pt]
        Control input saturation value ($\pm$) & 4.5e+1 & 3.0e-1 & 3.0e-1 & 1.6e-1 & 6.0e-2 \\[5pt]
        Number of time steps for each stair step (open loop signal) & 20 & 10 & 10 & 10 & 10 \\[5pt]
        Stair signal maximum amplitude & 4.5e+1 & 3.0e-1 & 3.0e-1 & 1.6e-1 & 6.0e-2 \\[5pt]
        maximum amplitude decay at each iteration & 6.5e-1 & 8.0e-1 & 8.0e-1 & 8.0e-1 & 9.0e-1 \\[5pt]
        Maximum number of points sampled for training NNSM & 4000 & 4000 & 4000 & 4000 & 4000 \\[5pt]
        NNSM learning rate & 6.0e-3 & 6.0e-3 & 6.0e-3 & 5.0e-3 & 1.0e-2 \\[5pt]
        Number of epochs at each training iteration for NNSM & 1000 & 1000 & 1000 & 2000 & 2000 \\[5pt]
        Number of neurons at NNSM hidden layers & [18] & [18] & [18] & [80] & [100, 80] \\[5pt]
        NNSM training set ratio (fraction of sampled data) & 7.0e-1 & 7.0e-1 & 7.0e-1 & 7.0e-1 & 7.0e-1 \\[5pt]
        Hidden layers L2 regularization & 1.0e-5 & 1.0e-5 & 1.0e-5 & 1.0e-5 & 1.0e-5 \\[5pt]
        Use sparsity layer & False & False & True & True & True \\[5pt]
        Sparsity layer L1 regularization & -- & -- & 1.0e-4 & 1.0e-5 & 1.0e-4 \\[5pt]
        Sparsity layer truncation tolerance & -- & -- & 3.0e-3 & 3.0e-3 & 3.0e-3 \\[5pt]
        Number of control inputs & 1 & 3 & 3 & 8 & 1 \\[5pt]
        Number of states & 3 & 60 & 60 & 664 & 306 \\[5pt]
        Maximum number of points sampled for training NNC & 1000 & 1000 & 1000 & 1000 & 1000 \\[5pt]
        NNC learning rate & 4.0e-3 & 1.0e-2 & 1.0e-2 & 1.0e-2 & 1.0e-2 \\[5pt]
        Number of epochs at each training step for NNC & 100 & 200 & 200 & 300 & 300 \\[5pt]
        Training finite horizon length & 10 & 50 & 50 & 140 & 140 \\[5pt]
        NNC training set ratio (fraction of sampled data) & 9.0e-1 & 9.0e-1 & 9.0e-1 & 9.0e-1 & 9.0e-1 \\[5pt]
        Number of neurons at NNC hidden layers & [8] & [8] & [8] & [8] & [8] \\[5pt]
        NNC control saturation & 4.5e+1 & 3.0e-1 & 3.0e-1 & 1.6e-1 & 6.0e-2 \\[5pt]
        Control input weight in loss function & 5.0e-3 & 5.0e-3 & 5.0e-3 & 1.0e-1 & 1.0e-1 \\[5pt]
        Initial guess for equilibrium point & [1.0e-1, 1.0e-1, 1.0e-1] & [1.0e-1, $\dots$, 1.0e-1] & [1.0e-1, $\dots$, 1.0e-1] & 1.0e-1 for x-velocity, 0.0e-0 for y-velocity & 1.0e-1 for x-velocity, 0.0e-0 for y-velocity\\[5pt]
        Number of steps for Newton method (equilibrium estimation) & 10 & 10 & 10 & 10 & 10 \\[5pt]
  \end{tabular}
}}
  \caption{Hyperparameters for each case studied}
  \label{tab:hp}
  \end{center}
\end{table}

\bibliographystyle{jfm}
\bibliography{jfm-instructions}

\end{document}